\documentclass[prl,aps,preprint,superscriptaddress,showpacs]{revtex4-1}
\usepackage[]{graphicx}
\usepackage[]{hyperref}
\usepackage[]{amsmath}
\usepackage{braket}
\usepackage{color}
\usepackage[labelfont=bf]{caption}
\begin{document}

\title{Progress Towards a Capacitively Mediated CNOT Between Two Charge Qubits in Si/SiGe}

\date{\today}
\author{E. R. MacQuarrie}
\affiliation{University of Wisconsin, Madison, WI 53706, USA}
\author{Samuel F. Neyens}
\affiliation{University of Wisconsin, Madison, WI 53706, USA}
\author{J. P. Dodson}
\affiliation{University of Wisconsin, Madison, WI 53706, USA}
\author{J. Corrigan}
\affiliation{University of Wisconsin, Madison, WI 53706, USA}
\author{Brandur Thorgrimsson}
\affiliation{University of Wisconsin, Madison, WI 53706, USA}
\author{Nathan Holman}
\affiliation{University of Wisconsin, Madison, WI 53706, USA}
\author{M. Palma}
\affiliation{University of Wisconsin, Madison, WI 53706, USA}
\author{L. F. Edge}
\affiliation{HRL Laboratories, LLC, 3011 Malibu Canyon Road, Malibu, CA 90265, USA}
\author{Mark Friesen}
\affiliation{University of Wisconsin, Madison, WI 53706, USA}
\author{S. N. Coppersmith}
\affiliation{University of Wisconsin, Madison, WI 53706, USA}
\affiliation{University of New South Wales, Sydney, Australia}
\author{M. A. Eriksson}
\affiliation{University of Wisconsin, Madison, WI 53706, USA}
\email{maeriksson@wisc.edu}

\begin{abstract}

Fast operations, an easily tunable Hamiltonian, and a straightforward two-qubit interaction make charge qubits a useful tool for benchmarking device performance and exploring two-qubit dynamics. Here, we tune a linear chain of four Si/SiGe quantum dots to host two double dot charge qubits. Using the capacitance between the double dots to mediate a strong two-qubit interaction, we simultaneously drive coherent transitions to generate correlations between the qubits. We then sequentially pulse the qubits to drive one qubit conditionally on the state of the other. We find that a conditional $\pi$-rotation can be driven in just $74$~ps with a modest fidelity demonstrating the possibility of two-qubit operations with a $13.5$~GHz clockspeed. 

\end{abstract}

\maketitle

\section{Introduction}

With charge, valley, and spin degrees of freedom, quantum dots in silicon are promising hosts of many different types of qubits. Using the electronic spin state as the logical basis has enabled high fidelity single-qubit operations~\cite{yoneda2018} and demonstrations of two-qubit gates~\cite{brunner2011,veldhorst2015,zajac2018,watson2018,huang2019,xue2019,msimmons2019,sigillito2019}. To date, two-qubit gates in Si quantum dots have been mediated by a spin-spin exchange interaction or by coupling via a superconducting resonator~\cite{borjans2020}. 

Alternatively, a capacitive interaction can be used to coherently couple neighboring double dot qubits using the electronic charge degree of freedom. Capacitive coupling has been used to perform fast two-qubit operations in charge qubits~\cite{li2015} and singlet-triplet qubits~\cite{shulman2012,nichol2017} in GaAs quantum dot devices. 

In Si-based quantum dot devices, a strong~\cite{zajac2016} and tunable~\cite{neyens2019} capacitive interaction between double dots has been demonstrated and used to perform qubit control conditionally on the state of a classical two level system~\cite{ward2016}. Here, we build on these results by using a capacitive interaction to measure correlated oscillations between two simultaneously-driven charge qubits. We then use this interaction to drive a fast ($74$~ps) conditional $\pi$-rotation. 

\section{Device Details}

\begin{figure}[ht]
\includegraphics[width=\linewidth]{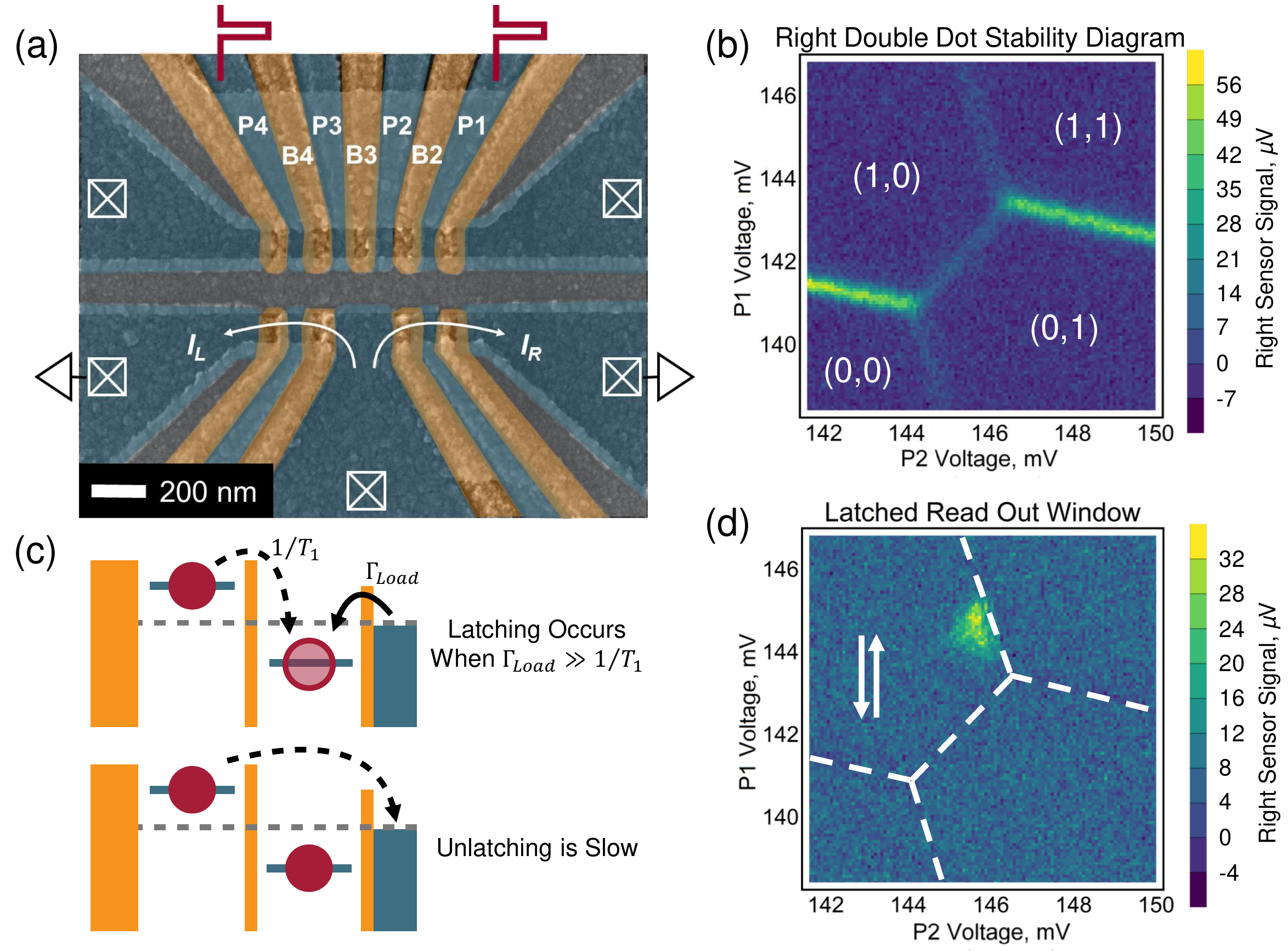} \\
\caption[fig:fig1] {(a) False-color scanning electron micrograph of a device nominally identical to the one measured here. Gates P1 and P4 are used for fast dc control. (b) Stability diagram of the right double dot tuned to the nominal (1,0)-(0,1) charge configuration. (c) Schematic depiction of latched state readout for the right double dot with a charge reservoir to the right and a hard-wall potential on the left. Latched readout projects the excited charge qubit state into a (1,1) charge configuration when the tunnel rate from the charge reservoir $\Gamma_{Load}$ exceeds the charge qubit relaxation rate $1/T_1$. (d) Stability diagram in (b) with fast dc control pulses applied to P1. The bright triangular region indicates the latched readout window, and the white arrows illustrate the applied dc pulse.}
\label{fig:fig1}
\end{figure}

To perform capacitively-coupled two-qubit measurements, we fabricate a linear chain of four quantum dots using an overlapping-aluminum gate architecture (Fig.~\ref{fig:fig1}a)~\cite{angus2007,borselli2015,zajac2016,neyens2019}. The fabrication details for this device have been reported in Ref.~\cite{neyens2019}. Measurements are performed in an Oxford Triton 400 dilution refrigerator with a $\sim15$~mK mixing chamber temperature. We tune the device to host two tunnel-coupled double dots, each nominally residing in the (1,0)-(0,1) charge configuration (Fig.~\ref{fig:fig1}b). In all measurements reported here, the double dot electron temperature was $T_0^{elec}=228$~mK and the charge reservoir temperature was $T_0^{res}=321$~mK~\cite{SI}. 

Two additional quantum dots are formed on the bottom half of the device to enable charge sensor readout of the qubit states. The current through the left charge sensor is measured with a room temperature current pre-amplifier, whereas the right charge sensor current is amplified at the mixing chamber of our dilution unit using a home-built two-stage cryogenic amplifier to enable high bandwidth readout~\cite{tracy2016}. While the right charge sensor only responds to the right double dot (RDD), during qubit operations the left charge sensor is able to sense both the RDD and the left double dot (LDD). As detailed in the Supplementary Information (SI), appropriate normalization measurements allow us to subtract the calibrated RDD signal from the left charge sensor data, enabling independent measurement of the two qubits~\cite{SI}. 

\section{Qubit Initialization, Control, and Latched Readout}

In our device, each double dot has an outer dot that neighbors a charge reservoir and an inner dot that is isolated from any reservoir. During qubit operations, we initialize each charge qubit at a large detuning $\varepsilon_I$ where one electron localizes into the qubit's outer dot ($\ket{\psi_0}=\ket{L}$ for the LDD; $\ket{\psi_0}=\ket{R}$ for the RDD). Single-qubit operations are described well by a charge qubit Hamiltonian
\begin{equation}
H_{1Q} = \frac{\varepsilon}{2}\sigma_z + t_c \sigma_x
\label{eq:1qHam}
\end{equation}
where $\varepsilon$ is the double dot detuning, $t_c$ is the tunnel coupling, and $\sigma_x$, $\sigma_z$ are the standard Pauli operators in the position basis $\{L,R\}$. 

To perform quantum control, we apply a fast dc pulse to move the system to $\varepsilon=0$. This rapid pulse non-adiabatically changes the qubit Hamiltonian to generate $\sigma_x$ rotations at a rate of $2t_c$. These rotations persist until the detuning is moved back to $\varepsilon_I$. If some fraction of the electron remains in the inner dot at the end of the coherent evolution, then there is nonzero probability that a second electron will tunnel from the reservoir into the outer dot before the qubit relaxes into $\ket{\psi_0}$. When this occurs, the qubit is projected into the (1,1) charge configuration and remains there until a co-tunneling process reinitializes the qubit into $\ket{\psi_0}$, providing a latched-state readout process~\cite{studenikin2012,harveycollard2018}. 

Using the metastable (1,1) charge configuration for latched-state readout provides two advantages. First, when the qubit enters the latched state, a second electron is added to the double dot system. This produces a larger shift in the charge sensor current than the mere relocation of a single electron. Secondly, this change in charge configuration persists for a much longer time because the co-tunneling process needed for reinitialization is generally much slower ($T_{Latch}>100$~ns) than charge qubit relaxation ($T_{1}<10$~ns in this device~\cite{SI}). Both of these mechanisms increase the signal generated by our qubit measurements. 

To maximize the probability that a driven state becomes latched, we tune the tunnel rate between the reservoir and the outer dot to be much larger than the charge relaxation rate between the two dots ($\Gamma_{Load}\gg 1/T_1$). Fig.~\ref{fig:fig1}c provides a schematic representation of this latched measurement strategy for the RDD, and Fig.~\ref{fig:fig1}d shows the latched state readout window that appears when dc pulses are applied to the stability diagram in Fig.~\ref{fig:fig1}b. This measurement was performed by shuttering our control pulses at a fixed repetition rate, locking in to the presence and absence of control pulses, and measuring the time-averaged charge sensor response. All qubit data reported here were measured with this latched-state, time-averaged technique. 

\section{Single-Qubit Measurements}

\begin{figure*}[ht]
\includegraphics[width=\linewidth]{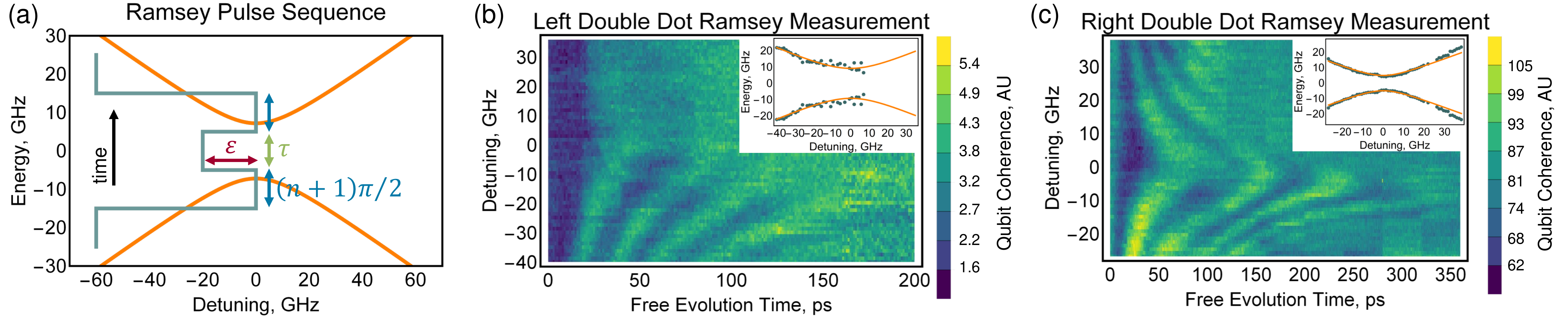} \\
\caption[fig:fig2] {(a) Dispersion and pulse sequence for measuring Ramsey oscillations at a detuning $\varepsilon$ for a free evolution time $\tau$. (b,c) Ramsey oscillations measured in the (b) left and (c) right double dots. The extracted charge qubit dispersions with (b) $t_c^L/h = 9.2$~GHz and (c) $t_c^R/h = 5.0$~GHz are shown in the insets. }
\label{fig:fig2}
\end{figure*}

With each qubit tuned to the nominal (1,0)-(0,1) charge configuration, we use dc control pulses to perform single-qubit Ramsey measurements of the qubit inhomogeneous dephasing times $T_2^*$. The pulse sequence (Fig.~\ref{fig:fig2}a) begins with initialization at large detuning $\varepsilon_I$. A sudden dc-shift to $\varepsilon=0$ turns on $\sigma_x$ rotations in the $\{L,R\}$ basis. After a $(n+1)\pi/2$ rotation, we apply a second dc-shift, moving to nonzero detuning and adding a $\sigma_z$ component to the Hamiltonian to start phase accumulation. Returning to $\varepsilon=0$ allows us to perform a second $(n+1)\pi/2$ rotation, projecting the accumulated phase onto the $z$-axis of the $\{L,R\}$ Bloch sphere. Finally, moving back to $\varepsilon_I$ for latched state readout maps the charge qubit coherence onto the measured charge sensor current~\cite{fujisawa2004}.

The Ramsey data for the LDD and RDD are shown in Fig.~\ref{fig:fig2}b,c, respectively. Both qubits display coherent behavior. By extracting the frequency of the Ramsey fringes as a function of detuning, we map the dispersion of our qubits and confirm that Eq.~\ref{eq:1qHam} appropriately describes each system. At large detuning ($\varepsilon/h>30$~GHz), however, the RDD dispersion begins to deviate from the expected charge qubit behavior. This could be due to timing artifacts in our control hardware as the rotation speed surpasses the $40$~ps rise time of our waveform generator. Alternatively, this could be evidence of a low-lying valley state generating additional curvature in the dispersion near $\varepsilon=0$~\cite{miValley2018}. 

The LDD Ramsey fringes lose all visibility for free evolution at $\varepsilon>0$. This could be due to imperfect pulse edges creating unintentional adiabaticity. Such an effect would have been more apparent in the LDD than in the RDD due to the larger tunnel coupling ($t_c^L/h = 9.2$~GHz versus $t_c^R/h = 5.0$~GHz) requiring faster rise times for true non-adiabatic control. 

For large detunings, the qubit dispersion is approximately linear in $\varepsilon$. Assuming non-Markovian detuning noise dominates the dephasing~\cite{dial2013}, we can fit the decaying coherence to a Gaussian envelope $e^{-t^2/T_2^{*2}}$ and extract the standard deviation of the quasistatic charge noise $\sigma_{\varepsilon} = h/\sqrt{2}\pi T_2^*$ where $h$ is Planck's constant. For the LDD and RDD, we find comparable values of $12.0\pm4.0$~$\mu$eV and $8.5\pm0.5$~$\mu$eV, respectively (additional details in the SI~\cite{SI}). 

\section{Correlated Oscillations}

\begin{figure}[ht]
\includegraphics[width=10cm]{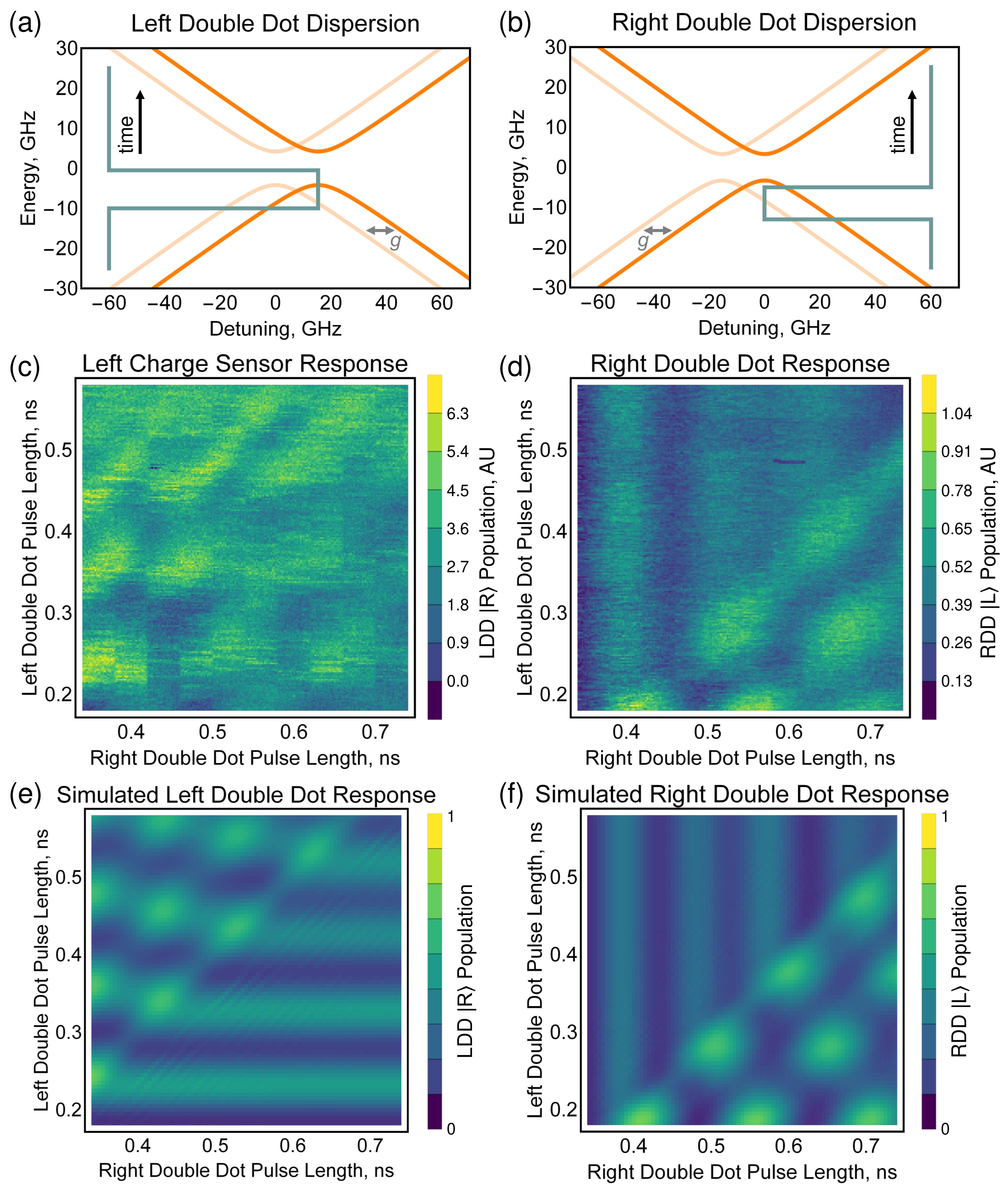} \\
\caption[fig:fig3] {(a,b) Dispersions and pulse sequences for simultaneously driving two charge qubits. (c,d) Measured two-qubit response to simultaneous driving. In (c), charge sensor crosstalk has been subtracted~\cite{SI}, and the black pixels lie outside the range of the plotted color scale. In (d), a jump in the charge sensor has been normalized out of the data~\cite{SI}. (e,f) Simulated two-qubit response to simultaneous driving. In this measurement, the right double dot pulse starts $150$~ps before the left double dot pulse. We note that the time evolution in this figure occurs near each qubit's anti-crossing, so coherent oscillations persist for longer times than those in Fig. 2. }
\label{fig:fig3}
\end{figure}

The two qubits in our device are capacitively coupled with a gate-voltage tunable coupling coefficient $g$~\cite{neyens2019}. In the two-qubit position basis $\{LL,LR,RL,RR\}$, the Hamiltonian describing this coupled system can be written as~\cite{li2015}
\begin{equation}
\begin{split}
H_{2Q} =& \frac{\varepsilon_L}{2}\sigma_z\otimes I + t_c^L \sigma_x\otimes I + \frac{\varepsilon_R}{2}I \otimes \sigma_z\\
&+t_c^R I\otimes \sigma_x +\frac{g}{4}(I-\sigma_z)\otimes(I-\sigma_z)
\end{split}
\label{eq:cqHam}
\end{equation}
where $\varepsilon_L$ ($\varepsilon_R$) and $t_c^L$ ($t_c^R$) are the detuning and tunnel coupling in the LDD (RDD) and $I$ is the identity matrix. The $\sigma_z\otimes\sigma_z$ nature of the capacitive interaction generates a detuning offset in one qubit conditionally on the state of the other (Figs.~\ref{fig:fig3}a,b). This capacitive interaction can be used to build state correlations between the two qubits.

In order to observe such correlations, we first tune a Hamiltonian with $t_c^L/h=4.2$, $t_c^R/h=3.3$, and $g/h=15.3$~GHz. We then simultaneously (modulo some fixed time offset) shift both qubits to their respective anti-crossings at $\varepsilon_L=g$ and $\varepsilon_R=0$ for times $\tau_L$ and $\tau_R$. At the end of $\tau_L$ ($\tau_R$), we shift the LDD (RDD) into its readout window for projection into the latched state. By independently varying $\tau_L$ and $\tau_R$ as shown in Figs.~\ref{fig:fig3}c,d, we can build up correlations during the simultaneous qubit evolutions, stop the rotations of one qubit, and observe the effect of those correlations in the continued evolution of the other qubit. Importantly, this measurement allows us to feedback on the time offset and sync our fast dc pulses at the mixing chamber to within $\sim80$~ps.

As shown in Fig.~\ref{fig:fig3}e,f, we recreate the measured two-qubit evolution by numerically solving the von Neumann equation using the Hamiltonian presented in Eq.~\ref{eq:cqHam}. Dephasing from charge noise is included by convolving this simulation with perturbations to both $\varepsilon_L$ and $\varepsilon_R$ (\textit{i.e.} $\varepsilon_i\rightarrow\varepsilon_i+\delta\varepsilon_i$). We assume these perturbations follow Gaussian distributions with standard deviations given by $\sigma_{\varepsilon}=12.0$ and $8.5$~$\mu$eV, respectively. Notably, the only free parameter in this simulation is the $150$~ps fixed offset between the rising edge of the two pulses. Additional simulation details are provided in the SI~\cite{SI}. 

\section{Towards a Two-Qubit Gate}

\begin{figure}[ht]
\includegraphics[width=\linewidth]{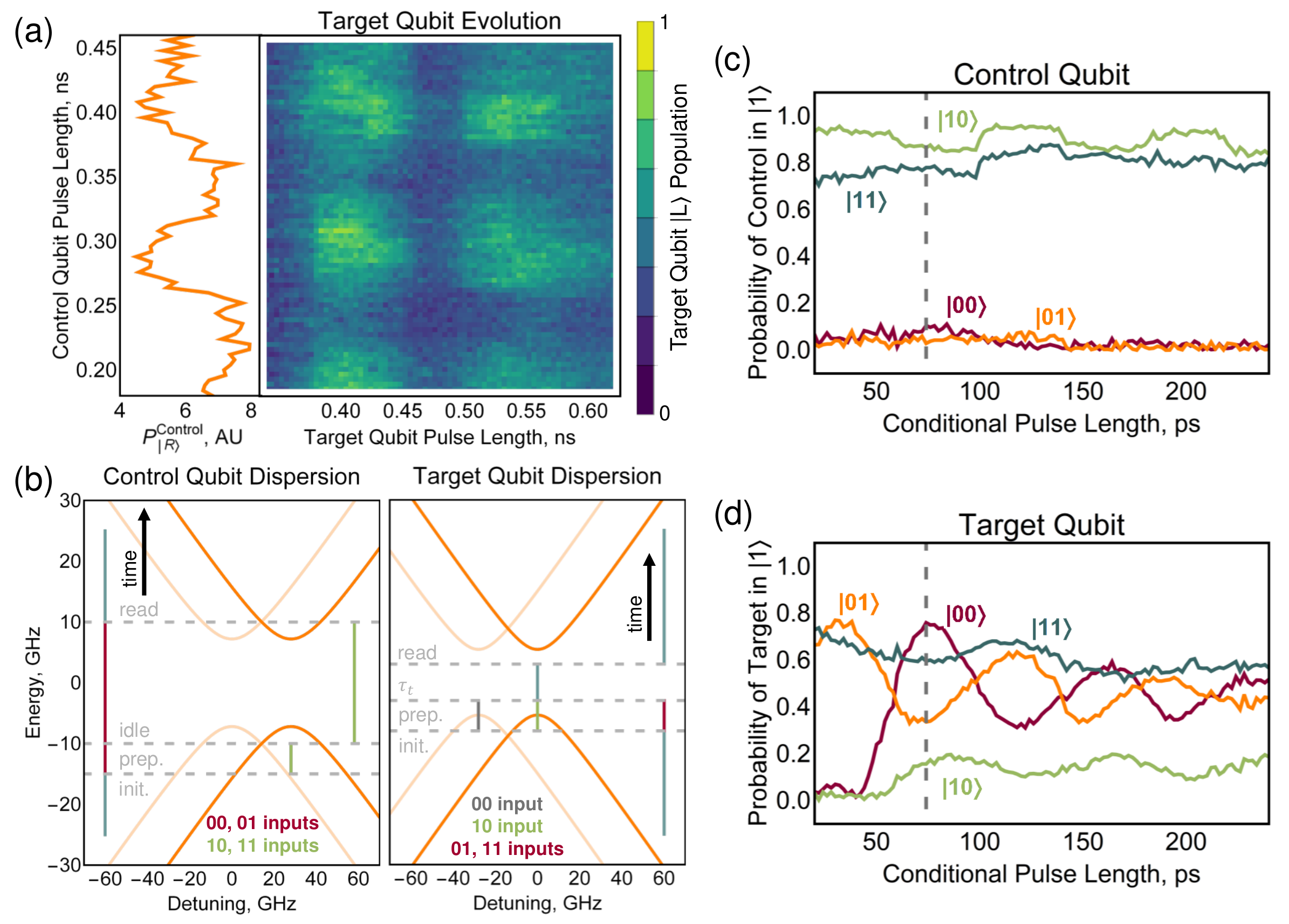} \\
\caption[fig:fig4] {(a) Evolution of the target qubit conditional on the coherent driving of the control qubit. A one-dimensional slice of the control qubit evolution is plotted on the left. (b) Dispersions and pulse sequences used to measure conditional rotations. The different control sequences used to prepare and measure the four input states are color-coded. (c,d) Evolution of the four input states in the (c) control and (d) target qubits. Charge sensor crosstalk has been subtracted from the control qubit data~\cite{SI}. }
\label{fig:fig4}
\end{figure}

The capacitive interaction can also be used to drive one qubit conditionally on the state of the other as has been demonstrated experimentally in GaAs charge qubits~\cite{li2015} and proposed theoretically in capacitively coupled Si/SiGe quantum dot hybrid qubits (QDHQs)~\cite{frees2018}. To demonstrate conditional rotations, we designate the LDD the control qubit and the RDD the target. We shift the control qubit to $\varepsilon_L=g$ (the location of the LDD anti-crossing for the initialized state $\ket{LR}$), allow it to evolve for some time $\tau_L$, and then shift it to a large detuning $\varepsilon_{idle}$ that lies outside the readout window to prevent projection into the latched state. While the control qubit is idling, the target qubit is pulsed to $\varepsilon_R=0$ and is conditionally driven dependent on the state populations of the control qubit. This pulse on the target qubit constitutes our conditional driving. Both qubits are then moved into their readout windows for latched-state measurement. Because latching of the control qubit projects its state, it is critical to maintain the control qubit at its idle point during the target qubit evolution. We note that the control qubit dephases during this idle period, but because dephasing does not alter qubit state populations, this does not affect the target qubit evolution. The results of this measurement generate the patchwork pattern shown in Fig.~\ref{fig:fig4}a, which is a hallmark of conditional evolution. 

Next, we characterize the fidelity of a conditional $\pi$-rotation at a tuning where $t_c^L/h=7.2$, $t_c^R/h=5.4$, and $g/h=28$~GHz~\cite{SI}. Our latched-state measurement technique provides the time-averaged values of $\langle\sigma_z\otimes I\rangle$ and $\langle I \otimes\sigma_z\rangle$. Without joint single shot readout or a verified, high fidelity two-qubit gate, this is not enough information to perform two-qubit tomography~\cite{vandersypen2005}, so we restrict our analysis to input states for which both qubits are expected to evolve into single-qubit eigenstates. For these inputs, we can assume the resulting two-qubit state is separable and our readout provides the appropriate populations for construction of the truth table $M_{exp}$ describing our conditional operation.

To measure $M_{exp}$, we define the initialized two-qubit state $\ket{LR}=\ket{00}$ and follow the pulse sequences shown in Fig.~\ref{fig:fig4}b to prepare each input state $\{CT\} = \{00,01,10,11\}$. We then measure the resulting output after application of an additional driving pulse of length $\tau_{t}$ on our target qubit. As discussed in the SI, the charge sensor dedicated to the control qubit measures both qubits simultaneously. To account for this, we use the calibrated signal from the target qubit's charge sensor to isolate the control qubit response. We then perform a maximum likelihood estimate to ensure positive probabilities~\cite{SI,james2001}. Fig.~\ref{fig:fig4}c,d show the results of this measurement. 

Selecting $\tau_{t}=74$~ps maximizes the average of the logical state input fidelities (the inquisition $\mathcal{I}$~\cite{white2007}) at a modest value of $\mathcal{I}=63\%$. At this point, in the $\{00,01,10,11\}$ basis,
\begin{equation}
M_{exp} = 
\begin{pmatrix} 
0.22 & 0.65 & 0.12 & 0.09 \\ 
0.68 & 0.33 & 0.02 & 0.13 \\
0.03 & 0.02 & 0.73 & 0.32 \\
0.08 & 0.01 & 0.13 & 0.46
\end{pmatrix} 
\end{equation}
which we compare to an ideal conditional $\pi$-rotation
\begin{equation}
M_{\pi} = 
\begin{pmatrix} 
0 & 1 & 0 & 0 \\ 
1 & 0 & 0 & 0 \\
0 & 0 & 1 & 0 \\
0 & 0 & 0 & 1
\end{pmatrix}.
\end{equation}
Notably, the input state that requires the most state preparation ($\ket{11}$) has a significantly lower fidelity ($46\%$) than the other input states. This suggests that state preparation errors are the dominant source of infidelity in our conditional operation. 

\section{Outlook}

Although the $74$~ps conditional $\pi$-rotation demonstrated here is consistent with a two-qubit CNOT, the dephasing of the control qubit during its idle step limits any claim of a coherent two-qubit processor. Nevertheless, the $13.5$~GHz two-qubit clockspeed highlights the benefit of using the strong capacitive interaction for inter-qubit coupling. Encoded qubits that have a tunable electric dipole moment such as the QDHQ stand to benefit from this fast gate speed without suffering from dephasing during idle periods. Compared to the charge qubits used in this work, higher fidelity single-qubit operations~\cite{kim2015} and longer coherence times~\cite{thorgrimsson2017} for the QDHQ could also reduce state preparation errors and enable the extended pulse sequences needed for a multi-qubit processor. 

In summary, we have demonstrated correlated and conditional evolution between two capacitively coupled charge qubits. After quantifying the single-qubit coherences, we simultaneously drove coherent rotations in both qubits to demonstrate correlated two-qubit evolution. We then operated in a sequential-driving mode to demonstrate a fast ($74$~ps) conditional $\pi$-rotation with a modest average fidelity ($63\%$) that was likely limited by state-preparation errors. These results represent an important proof of principle demonstration for fast two-qubit interactions in Si/SiGe double dot qubits. 

\section{Acknowledgments}

This research was sponsored in part by the Army Research Office (ARO) under Grant Number W911NF-17-1-0274 and by the Vannevar Bush Faculty Fellowship program under ONR grant number N00014-15-1-0029. We acknowledge the use of facilities supported by NSF through the UW-Madison MRSEC (DMR-1720415). The views and conclusions contained in this document are those of the authors and should not be interpreted as representing the official policies, either expressed or implied, of the Army Research Office (ARO), or the U.S. Government. The U.S. Government is authorized to reproduce and distribute reprints for Government purposes notwithstanding any copyright notation herein.

\section{Supplementary Information}

\subsection{Crosstalk Subtraction and Maximum Likelihood Estimation}

\begin{figure}[ht]
\includegraphics[width=\linewidth]{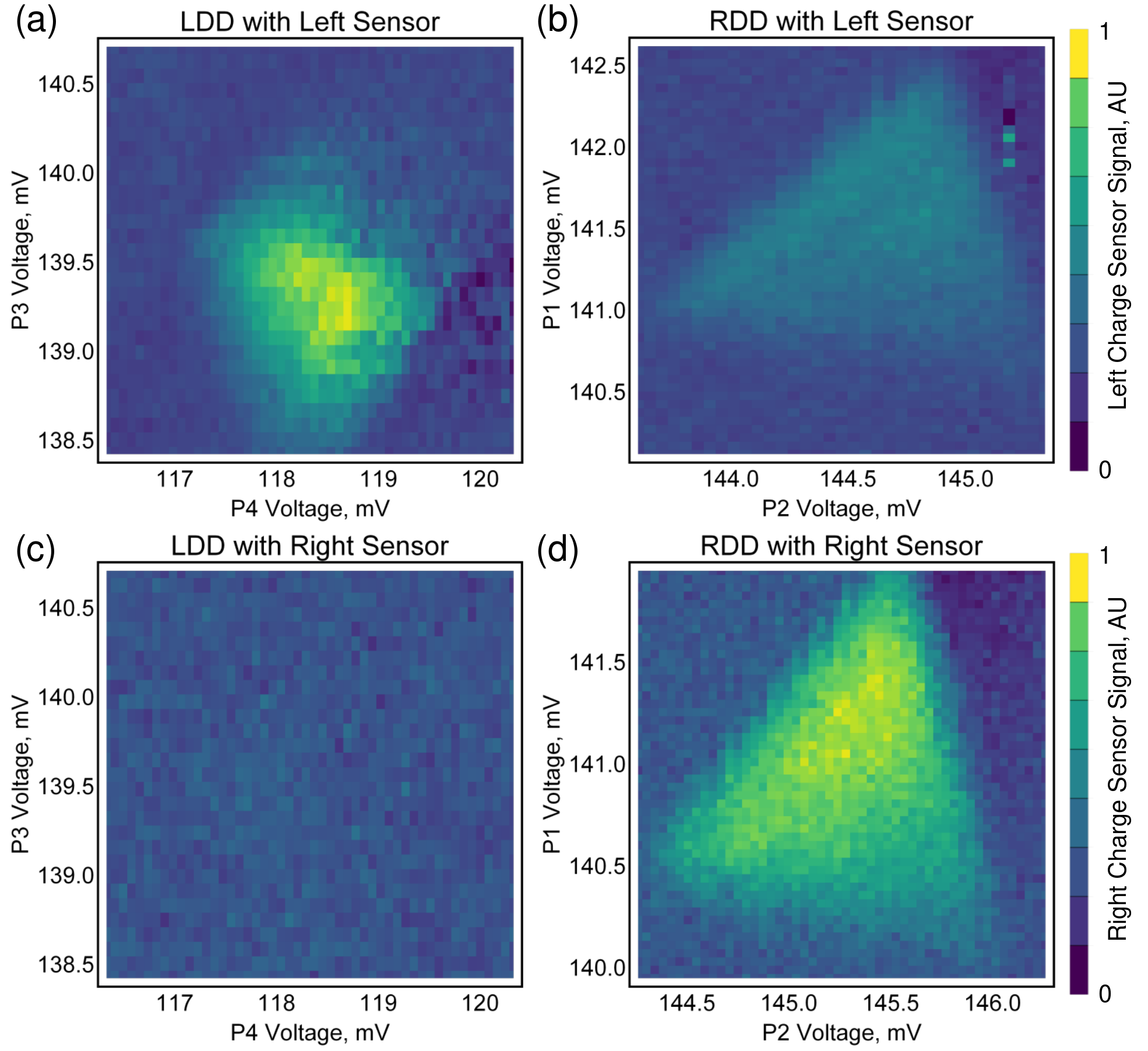} \\
\caption[fig:crosstalk1] {Measurements of the latched-state readout windows for the (a,c) LDD and (b,d) RDD using the (a,b) left charge sensor and (c,d) right charge sensor. }
\label{fig:crosstalk1}
\end{figure}

During our two-qubit measurements, the left charge sensor was sensitive to both the LDD and RDD qubits, whereas the right charge sensor was only sensitive to the RDD qubit dynamics. This crosstalk is demonstrated in Fig.~\ref{fig:crosstalk1}. With control pulses applied to the RDD, both the right and left charge sensors detect the RDD latched-state readout window. When pulses are applied to the LDD, however, only the left charge sensor measures the LDD latched-state readout window. This crosstalk obfuscates the LDD qubit dynamics, but appropriate normalization measurements allow us to deconvolve the LDD and RDD signal from the left sensor data. 

Since the right sensor only measures RDD qubit dynamics, two normalization measurements are performed for this signal. First, the right sensor is measured after the LDD and RDD qubits have been initialized into $\Ket{L}$ and $\Ket{R}$, respectively, providing $\mathcal{R}_{00}$. Next, the right sensor is measured after the RDD qubit has been pulsed into the (1,1) latched state. This is done by rapidly shifting the RDD to a large negative detuning, delaying at that point until the system has relaxed into $\ket{L}$, and rapidly shifting back to the readout window for latching. This second measurement provides $\mathcal{R}_{01}$. 

The left charge sensor measures both qubits, so more normalization measurements are required to deconvolve its signal. First, the pulses described in the previous paragraph are repeated, and the left sensor current is monitored. This provides the quantities $\mathcal{L}_{00}$ and $\mathcal{L}_{01}$. These same measurements are then repeated again with the pulses applied to the LDD qubit instead of the RDD qubit to obtain $\mathcal{L}_{00}$ (again) and $\mathcal{L}_{10}$. A final normalization measurement applies pulses to both the LDD and the RDD to obtain $\mathcal{L}_{11}$. 

It is worth noting that our time-averaged measurement technique integrates signal over the entire duty cycle of the pulse sequence. This pollutes our data with signal generated during the manipulation portion of the duty cycle. For all of our measurements, however, the manipulation time is many orders of magnitude shorter than the measurement time and any pollution is negligible. This effect is most significant for the measurements of $\mathcal{R}_{01}$, $\mathcal{L}_{01}$, $\mathcal{L}_{10}$, and $\mathcal{L}_{11}$ where the manipulation time rises to $\sim1$\% of the total duty cycle. 

After obtaining normalization data, qubit measurements are performed to obtain the uncalibrated signal $\mathcal{L}$ and $\mathcal{R}$. Because the right charge sensor only measures the RDD qubit, we can first calibrate $\mathcal{R}$ to obtain the probability the RDD qubit has ended its evolution in state $\ket{1}$:
\begin{equation}
P_{\ket{1}}^{RDD}=\frac{\mathcal{R}-\mathcal{R}_{00}}{\mathcal{R}_{01}-\mathcal{R}_{00}}.
\label{eq:rddPop}
\end{equation}

The left charge sensor measures both qubits simultaneously. To account for this, we first need to determine how the two qubit signals are combined in the charge sensor response. From Fig.~\ref{fig:crosstalk1}, we see that the LDD and the RDD both contribute \textit{positively} to the left sensor signal, and comparing normalization pulses, we find $\mathcal{L}_{10}>\mathcal{L}_{11}>\mathcal{L}_{01}>\mathcal{L}_{00}$. Making the assumption of monotonic contributions to the charge sensor signal, we explain this behavior with a LDD signal whose dynamic range depends on the state of the RDD. If the RDD is in $\ket{0}$ then the LDD signal ranges from $\mathcal{L}_{10}$ to $\mathcal{L}_{00}$, whereas with the RDD in $\ket{1}$, the LDD contribution ranges from $\mathcal{L}_{11}$ to $\mathcal{L}_{01}$. The RDD contribution, however, always ranges from $\left(\mathcal{L}_{01}-\mathcal{L}_{00}\right)$ to $0$.

To apply this model to our data, we approximate the combined signal by 
\begin{equation}
\mathcal{L} = \mathcal{L}_{LDD}+\mathcal{L}_{RDD}
\label{eq:lsSignal}
\end{equation}
where $\mathcal{L}_{RDD}$ ($\mathcal{L}_{LDD}$) is the RDD's (LDD's) contribution to the left sensor signal. The calibrated right sensor signal allows us to calculate 
\begin{equation}
\mathcal{L}_{RDD}=P_{\ket{1}}^{RDD}\times\left(\mathcal{L}_{01}-\mathcal{L}_{00}\right).
\label{eq:calRCont}
\end{equation}
Combining Eqs.~\ref{eq:lsSignal} and~\ref{eq:calRCont} and calibrating with our normalization data, we can then write the probability the LDD qubit has ended its evolution in state $\ket{1}$ as
\begin{equation}
P_{\ket{1}}^{LDD} = \frac{\mathcal{L}-P_{\ket{1}}^{RDD}\times\left(\mathcal{L}_{01}-\mathcal{L}_{00}\right)-c_{min}}{c_{max}-c_{min}}
\label{eq:ctSub}
\end{equation}
where
\begin{equation}
c_{min} = \mathcal{L}_{01}P_{\ket{1}}^{RDD} + \mathcal{L}_{00}\left(1-P_{\ket{1}}^{RDD}\right)
\end{equation}
and 
\begin{equation}
c_{max} = \mathcal{L}_{11}P_{\ket{1}}^{RDD} + \mathcal{L}_{10}\left(1-P_{\ket{1}}^{RDD}\right)
\end{equation}
define the state-dependent ranges of $\mathcal{L}_{LDD}$. Notably, applying this procedure to our normalization pulses returns the expected probabilities. 

\begin{figure}[ht]
\includegraphics[width=\linewidth]{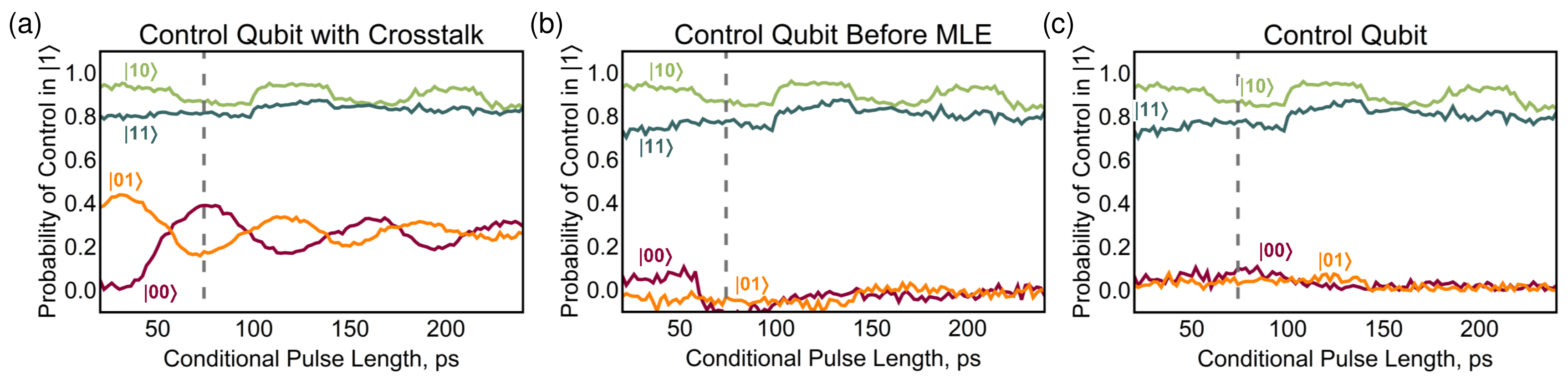} \\
\caption[fig:crosstalk2] {The control qubit data from our conditional measurements plotted (a) with and (b) without the target qubit crosstalk included. (c) The control qubit data after the maximum likelihood estimation has been performed. }
\label{fig:crosstalk2}
\end{figure}

The data shown in Fig.~4c of the main text is replotted in Figs.~\ref{fig:crosstalk2}a,b with and without the charge sensor crosstalk subtracted. For some portions of these data, this crosstalk removal procedure returns a negative probability (see Fig.~\ref{fig:crosstalk2}b). To make sense of this unphysical result, we apply a maximum likelihood estimator (MLE) to our single-qubit states to enforce positivity of the reported probabilities. Because we have assumed separable states in our conditional measurements, applying this MLE at the single-qubit or the two-qubit level provides identical results. 

\begin{figure}[ht]
\includegraphics[width=\linewidth]{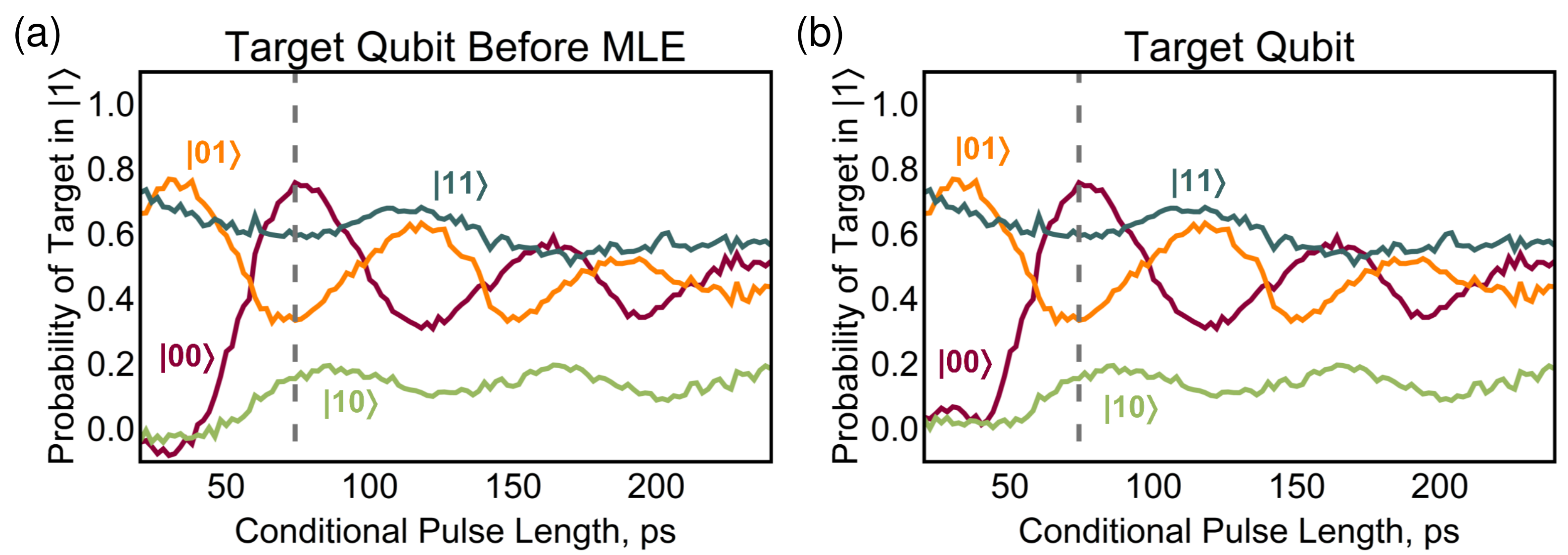} \\
\caption[fig:mleTarget] {The target qubit data from our conditional measurements plotted (a) before and (b) after the maximum likelihood estimation has been performed. }
\label{fig:mleTarget}
\end{figure}

The MLE aims to find the physically-valid density matrix $\rho_p$ that most closely approximates our measured density matrix $\rho_{exp}$. Since we can only measure the diagonal elements of $\rho_{exp}$, we adapt the MLE protocol used in Ref.~\cite{james2001} to neglect coherences. We constrain $\rho_p$ to be a non-negative, definite matrix by defining $\rho_p=\hat{T}^{\dagger}\hat{T}/\text{Tr}[\hat{T}^{\dagger}\hat{T}]$ where 
\begin{equation}
\hat{T} = 
\begin{pmatrix} 
t_1 & 0 \\ 
0 & t_2 
\end{pmatrix}.
\end{equation}
We then make the assumption that for each element $\rho_{exp,i}$ imperfections in our measurements generate a Gaussian probability of measuring the physical value $\rho_{p,i}$ and the standard deviation of that distribution is approximated by $\sqrt{\rho_{p,i}}$~\cite{james2001}. The probability that $\rho_{p}$ could produce $\rho_{exp}$ then becomes
\begin{equation}
P(\rho_{exp}) = \frac{1}{N}\prod_{i=1}^4 \text{exp}\left[\frac{-\left(\rho_{p,i}-\rho_{exp,i}\right)^2}{2\rho_{p,i}}\right]
\label{eq:mle1}
\end{equation}
where $N$ is a normalization constant. Rather than maximizing Eq.~\ref{eq:mle1}, we instead maximize its logarithm, which amounts to \textit{minimizing} the function
\begin{equation}
L(\rho_{exp}) = \sum_{i=1}^4 \frac{\left(\rho_{p,i}-\rho_{exp,i}\right)^2}{2\rho_{p,i}}.
\end{equation}
The diagonal elements of the resulting $\rho_p$ then fill in the columns of $M_{exp}$, providing the truth table quoted in the main text. The results of this MLE process are shown in Fig.~\ref{fig:crosstalk2}c for the control qubit and in Fig.~\ref{fig:mleTarget} for the target qubit data. 

\begin{figure}[ht]
\includegraphics[width=\linewidth]{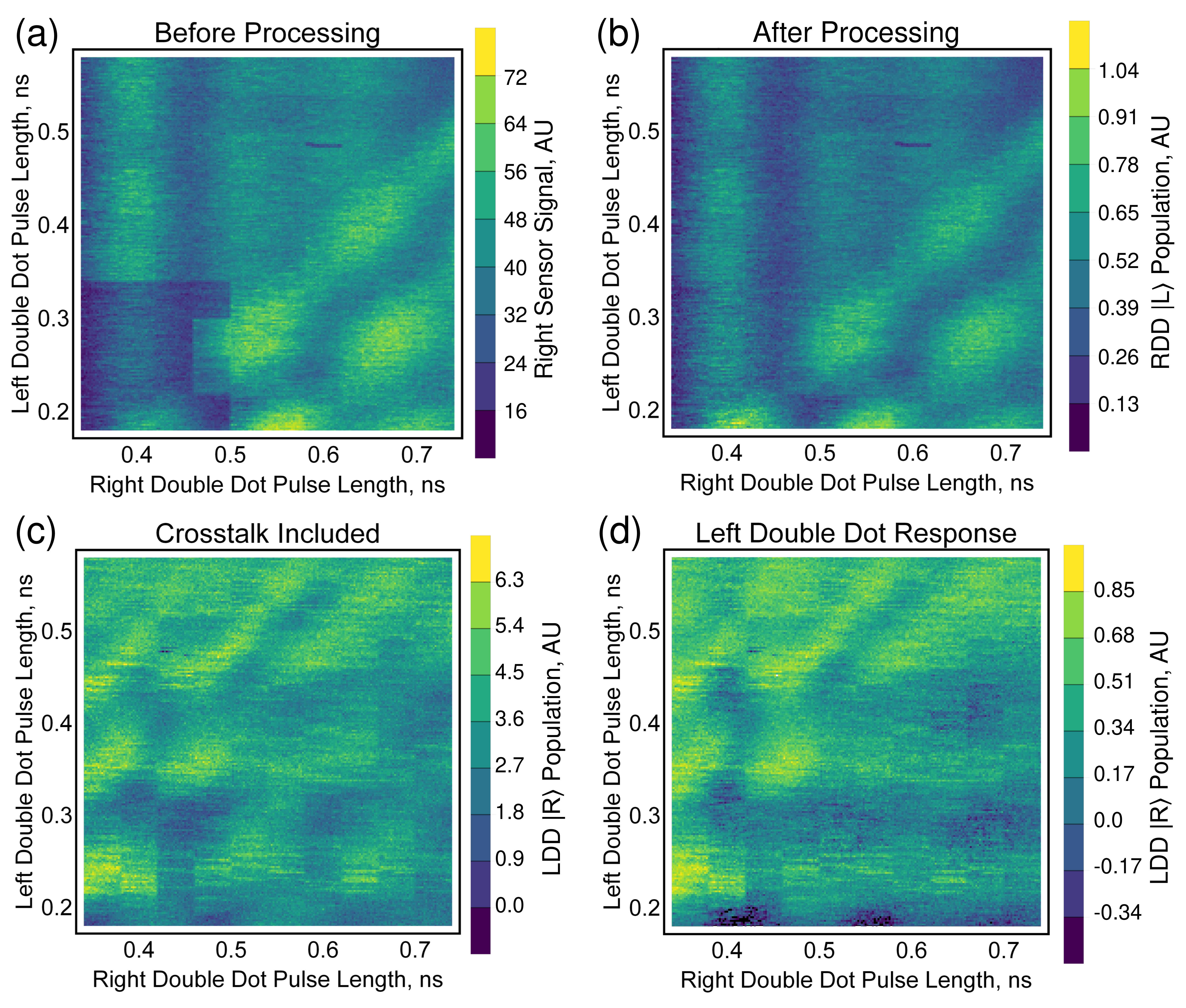} \\
\caption[fig:crosstalk3] {Correlated oscillation data for the (a,b) RDD and (c,d) LDD. (a,b) The RDD (a) before and (b) after the data has been processed to smooth a charge sensor jump. (c,d) The LDD data (c) before and (d) after the RDD crosstalk signal has been smoothed and removed. }
\label{fig:crosstalk3}
\end{figure}

For the data in Fig.~3 of the main text, we did not perform normalization measurements simultaneously with data acquisition. Moreover, the right charge sensor jumped during the course of the measurement. This jump created a discrete change in the charge sensor's dynamic range. To compensate for the effect of the jump, we split the data at the point of the jump and normalized each segment using the maximum and minimum values within that segment as approximations of $\mathcal{R}_{01}$ and $\mathcal{R}_{00}$, respectively. The effect of this procedure is demonstrated in Figs.~\ref{fig:crosstalk3}a,b. We then subtracted the RDD qubit signal from the left charge sensor data using the values of $\mathcal{L}_{01}$ and $\mathcal{L}_{11}$ measured during our conditional measurements and approximating $\mathcal{L}_{10}$ and $\mathcal{L}_{00}$ with the maximum and minimum values of the raw signal. The effect of this subtraction is shown in Fig.~\ref{fig:crosstalk3}c,d. Because we have approximated these normalization values, we plot the data with arbitrary units on the $z$-axis and do not apply the MLE for this measurement. 

\subsection{Fast Pulse Waveform Generation}

\begin{figure}[ht]
\includegraphics[width=\linewidth]{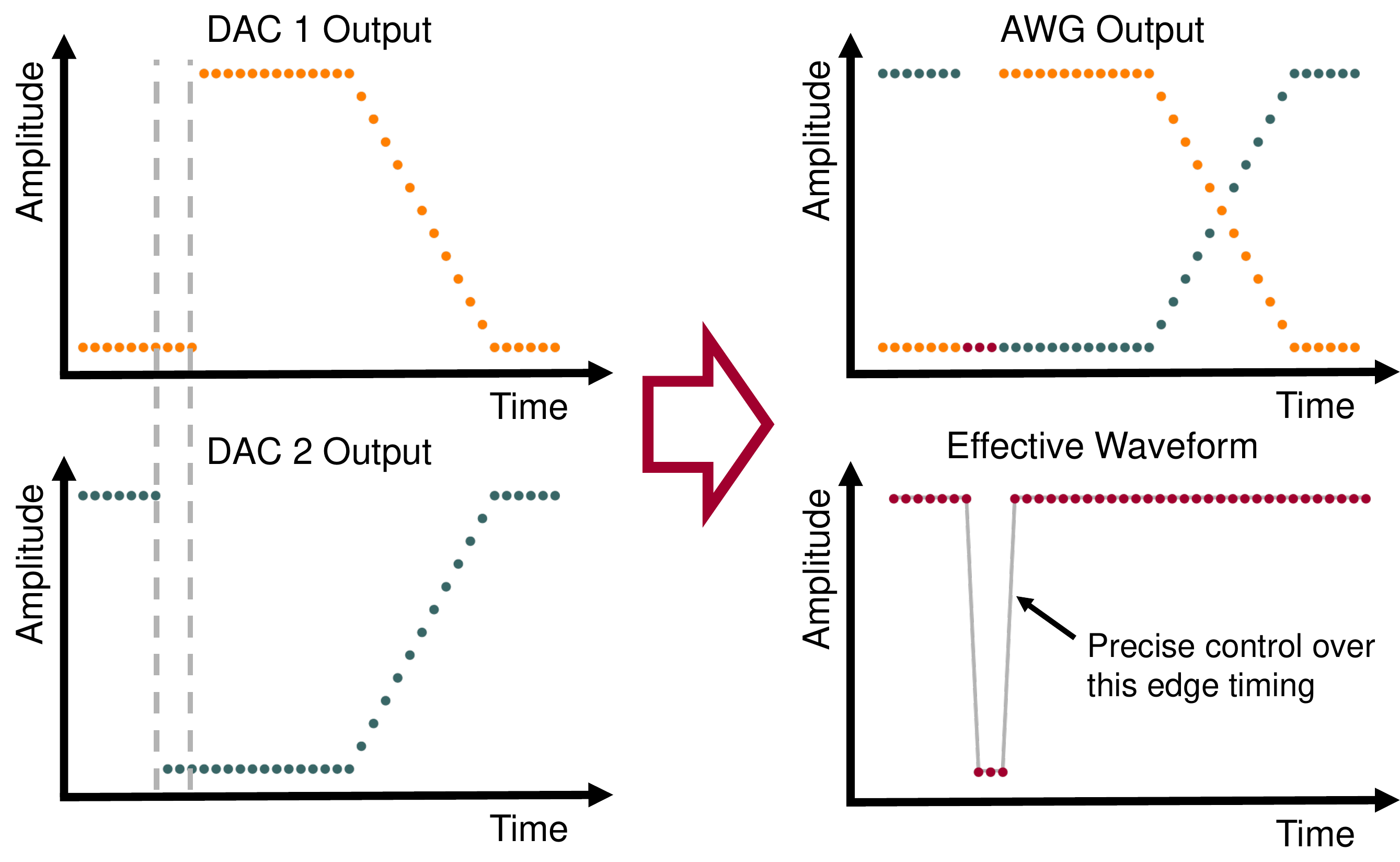} \\
\caption[fig:awg] {Schematic representation of our strategy for generating waveforms with $\sim$ps timing resolution}
\label{fig:awg}
\end{figure}

For each qubit, fast dc pulses were supplied by a Tektronix AWG 70001a. Internally, each waveform generator uses two interleaved $25$~GS/s digital-to-analog (DAC) converters to generate a $50$~GS/s waveform. We operate in a mode where, for a given AWG, each internal DAC outputs a distinct waveform. We output a positive waveform on one DAC and the negative of that same waveform plus some perturbation on the other. The internal power combiner of the AWG then sums the two waveforms, yielding just the perturbation, which we designate as our control pulse. We can control the phase delay between the two DACs with $\sim$ps resolution, providing precise control of the generated control pulse's duration. This method is depicted schematically in Fig.~\ref{fig:awg}.

For measurement sequences where multiple pulses were applied to the same qubit, this strategy of controlling the DAC phase delay only provides precise control over a single pulse edge in the sequence. Other pulses are constrained to durations that are multiples of the single DAC $40$~ps sampling resolution. For our conditional measurement (Fig.~4c,d of the main text), for instance, the target qubit input state preparation pulses were constrained to this $40$~ps grid. This pixelation likely contributed substantially to the state preparation errors that appear in our data. 

For two-qubit measurements, both AWGs were synced at the top of the fridge using a Tektronix Sync Hub. The uncorrected time delay between the two AWGs at the bottom of the fridge was measured to be $\sim0.75$~ns. 

\subsection{Electron Temperature}

We measured the electron temperature in the double dots (electron reservoirs) of our device by sweeping through a non-tunnel-broadened polarization line (charge transition) as a function of the mixing chamber temperature $T_{MC}$. For each temperature measurement, linecuts were collected at a range of $T_{MC}$ up to $350$~mK and then simultaneously fit to extract an effective electron temperature. Polarization lines were fit to a standard DiCarlo function with an electron temperature of $T_e=\sqrt{T_0^2+T_{MC}^2}$ where $T_0$ is the ideal electron temperature~\cite{dicarlo2004}. We note that this functional forms assumes an ideal charge qubit and thus ignores valley states which can lead to asymmetric lineshapes and/or modify the linewidths~\cite{hu2018}. Charge transitions to a reservoir were fit to a Fermi-Dirac distribution~\cite{maradan2014}. The voltage-to-energy lever arms were also free parameters in these fits but were constrained to be fixed across each linecut in a given dataset. 

 With this method, we obtained electron temperatures of $T_0=228\pm7$~mK for the RDD and $T_0=321\pm7$~mK for the left electron reservoir. These values are exceptionally high. We believe that these temperatures could be reduced in future experiments by improving the thermal anchoring of the dc lines at the mixing chamber. 
 
 \begin{figure}[ht]
\includegraphics[width=\linewidth]{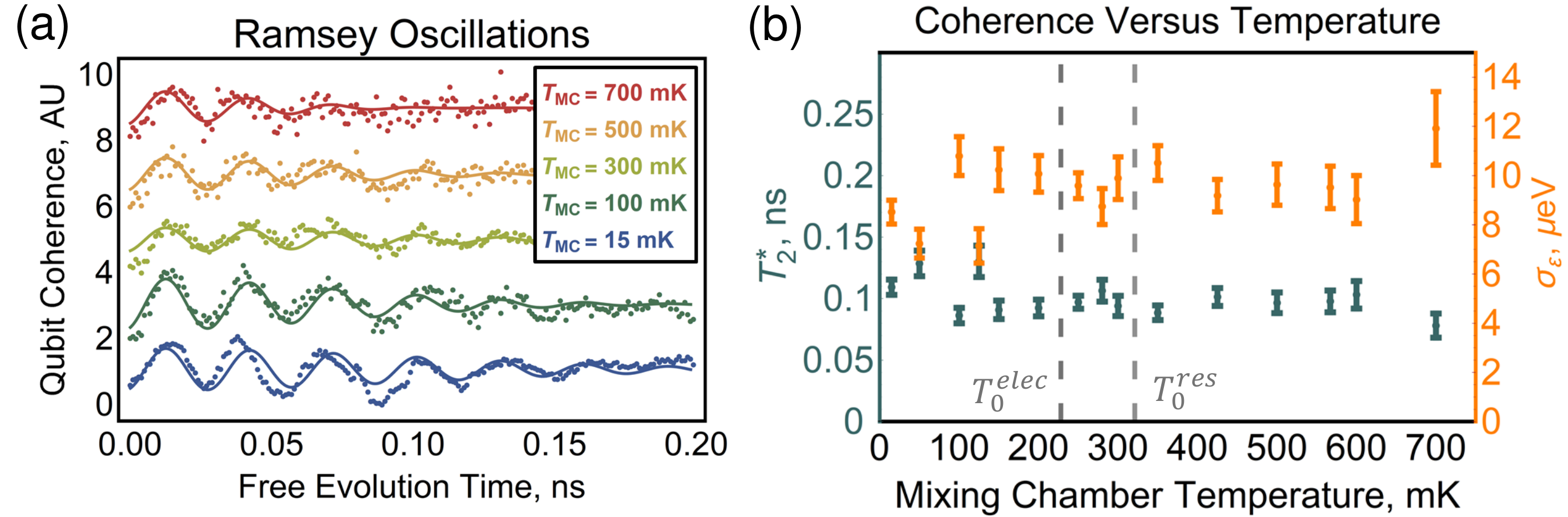} \\
\caption[fig:eTemp] {(a) Ramsey data measured on the right double dot as a function of temperature. These linecuts were taken at  $\varepsilon/h = -33$~GHz. (b) Inhomogeneous dephasing time and quasistatic charge noise extracted from the temperature-dependent Ramsey data.}
\label{fig:eTemp}
\end{figure}

To examine the prospect of operating our device at high temperatures, we measured Ramsey oscillations for the RDD as a function of the mixing chamber temperature (Fig.~\ref{fig:eTemp}a). Extracting $T_2^*$ and $\sigma_{\varepsilon}$ at each temperature (Fig.~\ref{fig:eTemp}b), we find that coherence persists up to $T_{MC}=700$~mK. In fact, these measurements were not limited by loss in coherence, but instead by a reduction in the visibility of our signal. At $700$~mK, the lifetime of our latched state had been reduced from $T_{Latch}\sim150$~ns to $T_{Latch}\sim40$~ns. Although not conclusive, these results are promising for the prospect of operating qubits with a charge-like degree of freedom at higher temperatures. 
 

\subsection{$T_{1}$ Measurements}

\begin{figure}[ht]
\includegraphics[width=\linewidth]{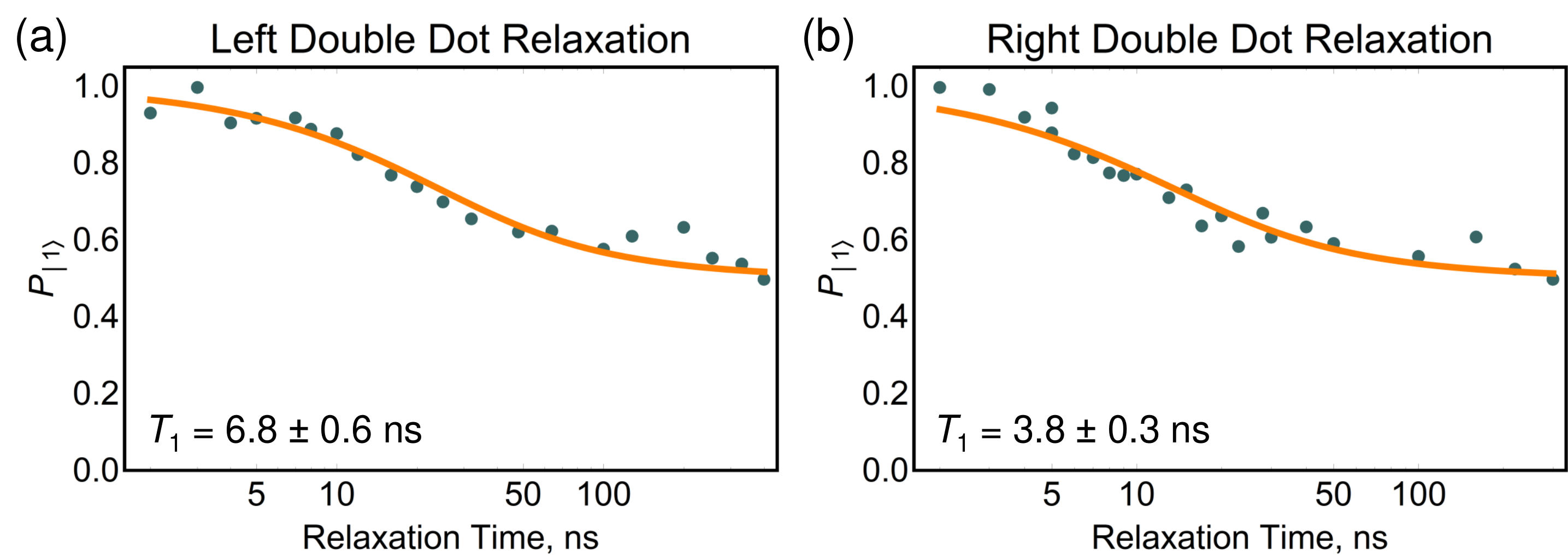} \\
\caption[fig:t1] {$T_1$ measurements for the (a) left double dot and (b) right double dot.}
\label{fig:t1}
\end{figure}

Following the method described in Ref.~\cite{wang2013}, we measured the relaxation time $T_1$ of our two charge qubits. For both qubits, we measured $T_1<10$~ns (Fig.~\ref{fig:t1}), which is short enough to prohibit ac driving of our charge qubits~\cite{dohun2015}. We speculate that this short relaxation time stems from increased electron-phonon scattering due to our high electron temperature. 

\subsection{Fitting Ramsey Data}

To extract the inhomogeneous dephasing time $T_2^*$, we neglect any valley or spin degrees of freedom and fit the charge qubit coherence $\rho_{LR}$ to the function
\begin{equation}
\rho_{LR} = A e^{-\tau^2/T_2^{*2}} \text{cos}\left(\omega \tau+\phi\right)+B
\label{eq:ramsey}
\end{equation}
where $A$ and $B$ are constants, $\tau$ is the free evolution time, $\omega$ is the qubit frequency at a given detuning, and $\phi$ is a fixed phase offset. 

To extract the charge qubit dispersions shown in the insets of Fig.~2a,b of the main text, we fit linecuts of the data to Eq.~\ref{eq:ramsey}. For the LDD data, the Ramsey fringe visibility vanishes for $\varepsilon>0$. The background level also drifts with the free evolution time $\tau$ in these data. To correct for this, we average all linecuts with $\varepsilon/h>23.4$~GHz where the fringe visibility has vanished and subtract this mean from the rest of the data before fitting. The $\sigma_{\varepsilon}=12.0\pm4.0$~$\mu$eV value for the LDD quasistatic charge noise was obtained by averaging the $T_2^*$ values returned from the fits for all $\varepsilon/h<-29.7$~GHz at which point $|\partial\omega/\partial\varepsilon|>0.85$.

When fitting the Ramsey measurements performed as a function of temperature (Fig.~\ref{fig:eTemp} in the SI), we fix $\phi$ and $\omega$ at each detuning to be the same for every temperature. The data in Fig.~\ref{fig:eTemp}b are the average results for linecuts in the detuning range $\varepsilon/h\in(-33,-29)$~GHz. The $\sigma_{\varepsilon}=8.5\pm0.5$~$\mu$eV value of the RDD quasistatic charge noise quoted in the main text was extracted from the $T_{MC}=15$~mK datum in this measurement. 

\subsection{Simulations of Correlated Oscillations}

\begin{figure}[ht]
\includegraphics[width=\linewidth]{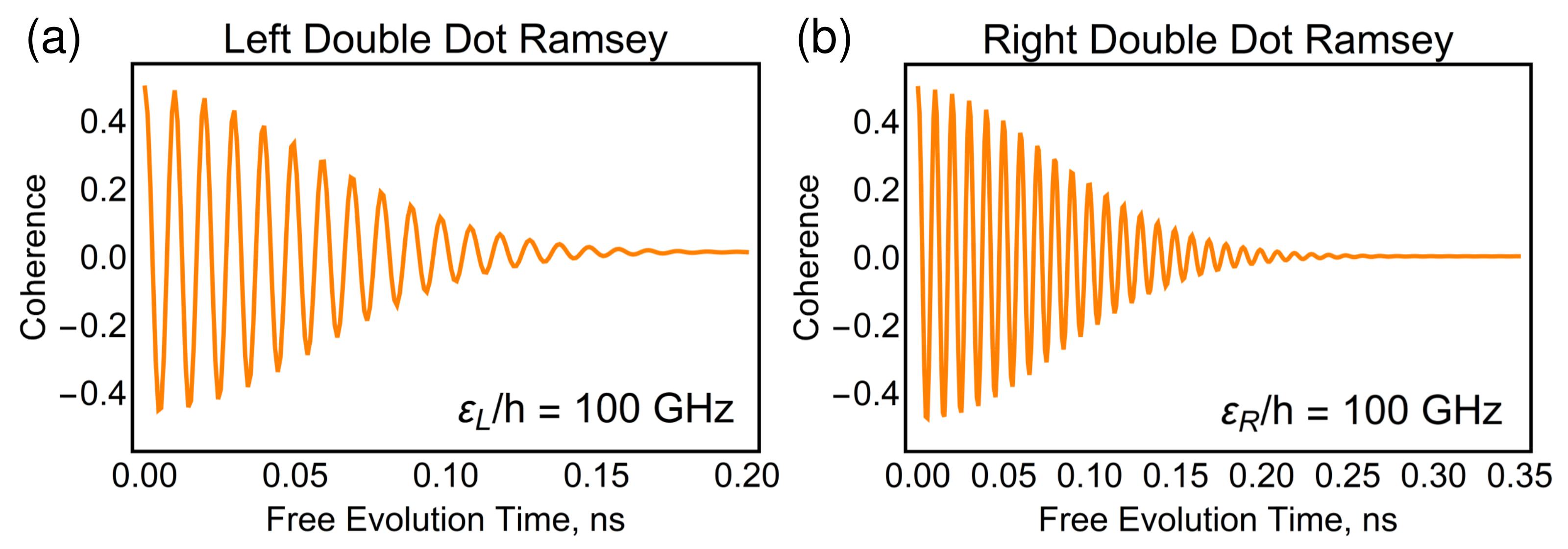} \\
\caption[fig:simRamsey] {Simulations of single qubit dephasing for the (a) left double dot and (b) right double dot.}
\label{fig:simRamsey}
\end{figure}

The simulation results presented in Fig.~3e,f of the main text were obtained by numerically solving the von Neumann equation 
\begin{equation}
i \hbar \frac{\partial\rho}{\partial t}=[H_{2Q},\rho].
\end{equation}
Here, $\hbar$ is the reduced Planck constant, $\rho$ is the density matrix for the two-qubit system, and $H_{2Q}$ is the Hamiltonian presented in Eq.~2 of the main text. Dephasing was included in the simulation by adding a perturbation to each double dot's detuning ($\varepsilon_i\rightarrow\varepsilon_i+\delta\varepsilon_i$), convolving the simulation with Gaussian distributions of $d\varepsilon_L$ and $d\varepsilon_R$, and normalizing appropriately. To verify the simulation reproduced the experimentally-measured coherence times, we simulated single qubit dephasing measurements in the large-detuning regime (Fig.~\ref{fig:simRamsey}). 

\subsection{Capacitive Shift of Latched State}

\begin{figure}[ht]
\includegraphics[width=\linewidth]{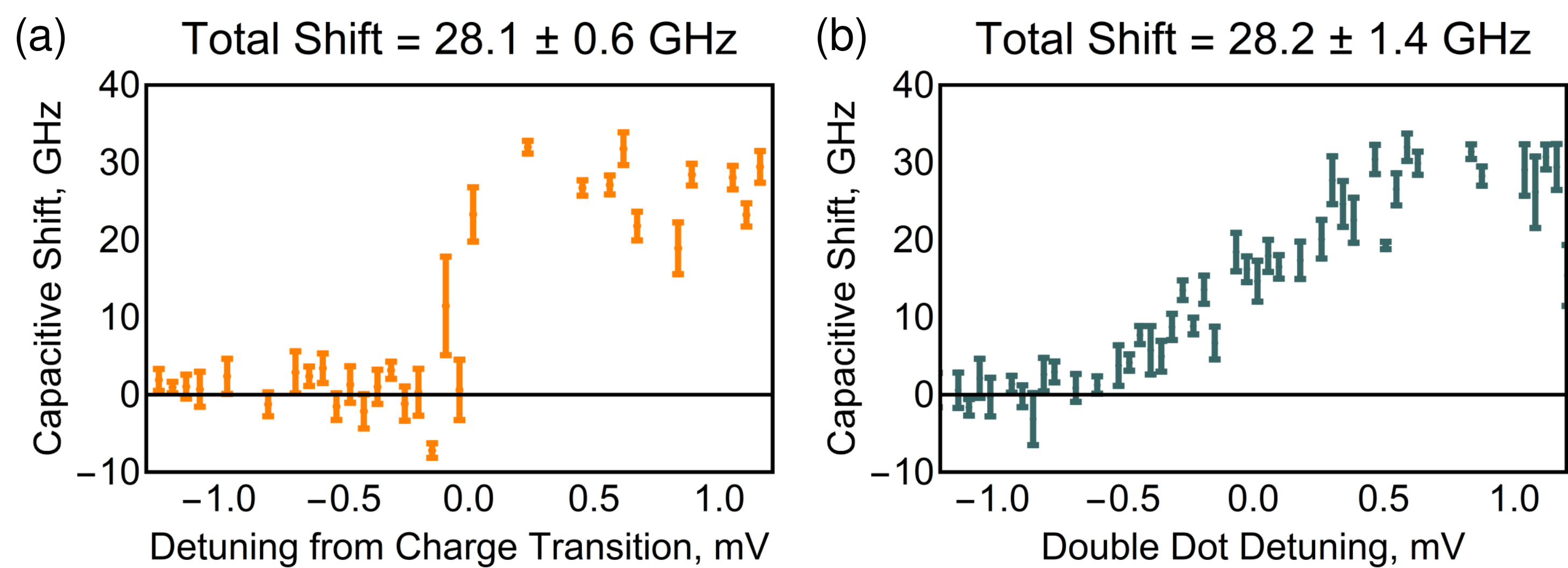} \\
\caption[fig:latchCap] {Capacitive shift experienced by the right double dot due to a transition in the left double dot from (a) the (1,0) to the (1,1) charge state and (b) the (1,0) to the (0,1) charge state. Note that the abrupt transition in (a) is because the charge transition being swept is a cotunneling process with a very slow tunnel rate~\cite{neyens2019}. }
\label{fig:latchCap}
\end{figure}

As discussed in the main text, shelving one double dot into its metastable latched state produces a capacitive shift in the other double dot. For our conditional measurements, we move the control qubit to an idle point during target qubit operations. This delays projection into the latched state until after our conditional rotation is complete, ensuring that any conditional behavior we detect results from the capacitive interaction between the two qubits.

Our measurement of correlated oscillations in Fig.~3 of the main text did not use idle points to delay projection into the latched state. This means that once we deviate from the diagonal that defines synchronized pulse tails, one qubit has been moved into its readout window and might have been projected into its latched state. However, extending the classical capacitance network model described in Ref.~\cite{neyens2019}, we can show that to first order in interdot capacitances the double dot capacitive shift $g$ is equal to the capacitive shift from the latched state $g_{Latch}$. For the simulation shown in Figs.~3e,f, we therefore use $g=g_{Latch}=15.3\times h$~GHz. This relation between $g$ and $g_{Latch}$ was verified via electrostatic measurements at the device tuning used for our conditional measurements (Fig.~\ref{fig:latchCap}) and is expected to hold at the tuning used in Fig.~3 of the main text.  

\section{References}
\bibliography{bibDots}

\begin{thebibliography}{36}%
\makeatletter
\providecommand \@ifxundefined [1]{%
 \@ifx{#1\undefined}
}%
\providecommand \@ifnum [1]{%
 \ifnum #1\expandafter \@firstoftwo
 \else \expandafter \@secondoftwo
 \fi
}%
\providecommand \@ifx [1]{%
 \ifx #1\expandafter \@firstoftwo
 \else \expandafter \@secondoftwo
 \fi
}%
\providecommand \natexlab [1]{#1}%
\providecommand \enquote  [1]{``#1''}%
\providecommand \bibnamefont  [1]{#1}%
\providecommand \bibfnamefont [1]{#1}%
\providecommand \citenamefont [1]{#1}%
\providecommand \href@noop [0]{\@secondoftwo}%
\providecommand \href [0]{\begingroup \@sanitize@url \@href}%
\providecommand \@href[1]{\@@startlink{#1}\@@href}%
\providecommand \@@href[1]{\endgroup#1\@@endlink}%
\providecommand \@sanitize@url [0]{\catcode `\\12\catcode `\$12\catcode
  `\&12\catcode `\#12\catcode `\^12\catcode `\_12\catcode `\%12\relax}%
\providecommand \@@startlink[1]{}%
\providecommand \@@endlink[0]{}%
\providecommand \url  [0]{\begingroup\@sanitize@url \@url }%
\providecommand \@url [1]{\endgroup\@href {#1}{\urlprefix }}%
\providecommand \urlprefix  [0]{URL }%
\providecommand \Eprint [0]{\href }%
\providecommand \doibase [0]{http://dx.doi.org/}%
\providecommand \selectlanguage [0]{\@gobble}%
\providecommand \bibinfo  [0]{\@secondoftwo}%
\providecommand \bibfield  [0]{\@secondoftwo}%
\providecommand \translation [1]{[#1]}%
\providecommand \BibitemOpen [0]{}%
\providecommand \bibitemStop [0]{}%
\providecommand \bibitemNoStop [0]{.\EOS\space}%
\providecommand \EOS [0]{\spacefactor3000\relax}%
\providecommand \BibitemShut  [1]{\csname bibitem#1\endcsname}%
\let\auto@bib@innerbib\@empty
\bibitem [{\citenamefont {Yoneda}\ \emph {et~al.}(2018)\citenamefont {Yoneda},
  \citenamefont {Takeda}, \citenamefont {Otsuka}, \citenamefont {Nakajima},
  \citenamefont {Delbecq}, \citenamefont {Allison}, \citenamefont {Honda},
  \citenamefont {Kodera}, \citenamefont {Oda}, \citenamefont {Hoshi},
  \citenamefont {Usami}, \citenamefont {Itoh},\ and\ \citenamefont
  {Tarucha}}]{yoneda2018}%
  \BibitemOpen
  \bibfield  {author} {\bibinfo {author} {\bibfnamefont {J.}~\bibnamefont
  {Yoneda}}, \bibinfo {author} {\bibfnamefont {K.}~\bibnamefont {Takeda}},
  \bibinfo {author} {\bibfnamefont {T.}~\bibnamefont {Otsuka}}, \bibinfo
  {author} {\bibfnamefont {T.}~\bibnamefont {Nakajima}}, \bibinfo {author}
  {\bibfnamefont {M.~R.}\ \bibnamefont {Delbecq}}, \bibinfo {author}
  {\bibfnamefont {G.}~\bibnamefont {Allison}}, \bibinfo {author} {\bibfnamefont
  {T.}~\bibnamefont {Honda}}, \bibinfo {author} {\bibfnamefont
  {T.}~\bibnamefont {Kodera}}, \bibinfo {author} {\bibfnamefont
  {S.}~\bibnamefont {Oda}}, \bibinfo {author} {\bibfnamefont {Y.}~\bibnamefont
  {Hoshi}}, \bibinfo {author} {\bibfnamefont {N.}~\bibnamefont {Usami}},
  \bibinfo {author} {\bibfnamefont {K.~M.}\ \bibnamefont {Itoh}}, \ and\
  \bibinfo {author} {\bibfnamefont {S.}~\bibnamefont {Tarucha}},\ }\href
  {\doibase 10.1038/s41565-017-0014-x} {\bibfield  {journal} {\bibinfo
  {journal} {Nature Nanotech.}\ }\textbf {\bibinfo {volume} {13}},\ \bibinfo
  {pages} {102} (\bibinfo {year} {2018})}\BibitemShut {NoStop}%
\bibitem [{\citenamefont {Brunner}\ \emph {et~al.}(2011)\citenamefont
  {Brunner}, \citenamefont {Shin}, \citenamefont {Obata}, \citenamefont
  {Pioro-Ladri\`ere}, \citenamefont {Kubo}, \citenamefont {Yoshida},
  \citenamefont {Taniyama}, \citenamefont {Tokura},\ and\ \citenamefont
  {Tarucha}}]{brunner2011}%
  \BibitemOpen
  \bibfield  {author} {\bibinfo {author} {\bibfnamefont {R.}~\bibnamefont
  {Brunner}}, \bibinfo {author} {\bibfnamefont {Y.-S.}\ \bibnamefont {Shin}},
  \bibinfo {author} {\bibfnamefont {T.}~\bibnamefont {Obata}}, \bibinfo
  {author} {\bibfnamefont {M.}~\bibnamefont {Pioro-Ladri\`ere}}, \bibinfo
  {author} {\bibfnamefont {T.}~\bibnamefont {Kubo}}, \bibinfo {author}
  {\bibfnamefont {K.}~\bibnamefont {Yoshida}}, \bibinfo {author} {\bibfnamefont
  {T.}~\bibnamefont {Taniyama}}, \bibinfo {author} {\bibfnamefont
  {Y.}~\bibnamefont {Tokura}}, \ and\ \bibinfo {author} {\bibfnamefont
  {S.}~\bibnamefont {Tarucha}},\ }\href {\doibase
  10.1103/PhysRevLett.107.146801} {\bibfield  {journal} {\bibinfo  {journal}
  {Phys. Rev. Lett.}\ }\textbf {\bibinfo {volume} {107}},\ \bibinfo {pages}
  {146801} (\bibinfo {year} {2011})}\BibitemShut {NoStop}%
\bibitem [{\citenamefont {Veldhorst}\ \emph {et~al.}(2015)\citenamefont
  {Veldhorst}, \citenamefont {Yang}, \citenamefont {Hwang}, \citenamefont
  {Huang}, \citenamefont {Dehollain}, \citenamefont {Muhonen}, \citenamefont
  {Simmons}, \citenamefont {Laucht}, \citenamefont {Hudson}, \citenamefont
  {Itoh}, \citenamefont {Morello},\ and\ \citenamefont
  {Dzurak}}]{veldhorst2015}%
  \BibitemOpen
  \bibfield  {author} {\bibinfo {author} {\bibfnamefont {M.}~\bibnamefont
  {Veldhorst}}, \bibinfo {author} {\bibfnamefont {C.~H.}\ \bibnamefont {Yang}},
  \bibinfo {author} {\bibfnamefont {J.~C.~C.}\ \bibnamefont {Hwang}}, \bibinfo
  {author} {\bibfnamefont {W.}~\bibnamefont {Huang}}, \bibinfo {author}
  {\bibfnamefont {J.~P.}\ \bibnamefont {Dehollain}}, \bibinfo {author}
  {\bibfnamefont {J.~T.}\ \bibnamefont {Muhonen}}, \bibinfo {author}
  {\bibfnamefont {S.}~\bibnamefont {Simmons}}, \bibinfo {author} {\bibfnamefont
  {A.}~\bibnamefont {Laucht}}, \bibinfo {author} {\bibfnamefont {F.~E.}\
  \bibnamefont {Hudson}}, \bibinfo {author} {\bibfnamefont {K.~M.}\
  \bibnamefont {Itoh}}, \bibinfo {author} {\bibfnamefont {A.}~\bibnamefont
  {Morello}}, \ and\ \bibinfo {author} {\bibfnamefont {A.~S.}\ \bibnamefont
  {Dzurak}},\ }\href {\doibase 10.1038/nature15263} {\bibfield  {journal}
  {\bibinfo  {journal} {Nature}\ }\textbf {\bibinfo {volume} {526}},\ \bibinfo
  {pages} {410} (\bibinfo {year} {2015})}\BibitemShut {NoStop}%
\bibitem [{\citenamefont {Zajac}\ \emph {et~al.}(2018)\citenamefont {Zajac},
  \citenamefont {Sigillito}, \citenamefont {Russ}, \citenamefont {Borjans},
  \citenamefont {Taylor}, \citenamefont {Burkard},\ and\ \citenamefont
  {Petta}}]{zajac2018}%
  \BibitemOpen
  \bibfield  {author} {\bibinfo {author} {\bibfnamefont {D.~M.}\ \bibnamefont
  {Zajac}}, \bibinfo {author} {\bibfnamefont {A.~J.}\ \bibnamefont
  {Sigillito}}, \bibinfo {author} {\bibfnamefont {M.}~\bibnamefont {Russ}},
  \bibinfo {author} {\bibfnamefont {F.}~\bibnamefont {Borjans}}, \bibinfo
  {author} {\bibfnamefont {J.~M.}\ \bibnamefont {Taylor}}, \bibinfo {author}
  {\bibfnamefont {G.}~\bibnamefont {Burkard}}, \ and\ \bibinfo {author}
  {\bibfnamefont {J.~R.}\ \bibnamefont {Petta}},\ }\href {\doibase
  10.1126/science.aao5965} {\bibfield  {journal} {\bibinfo  {journal}
  {Science}\ }\textbf {\bibinfo {volume} {359}},\ \bibinfo {pages} {439}
  (\bibinfo {year} {2018})}\BibitemShut {NoStop}%
\bibitem [{\citenamefont {Watson}\ \emph {et~al.}(2018)\citenamefont {Watson},
  \citenamefont {Philips}, \citenamefont {Kawakami}, \citenamefont {Ward},
  \citenamefont {Scarlino}, \citenamefont {Veldhorst}, \citenamefont {Savage},
  \citenamefont {Lagally}, \citenamefont {Friesen}, \citenamefont
  {Coppersmith}, \citenamefont {Eriksson},\ and\ \citenamefont
  {Vandersypen}}]{watson2018}%
  \BibitemOpen
  \bibfield  {author} {\bibinfo {author} {\bibfnamefont {T.~F.}\ \bibnamefont
  {Watson}}, \bibinfo {author} {\bibfnamefont {S.~G.~J.}\ \bibnamefont
  {Philips}}, \bibinfo {author} {\bibfnamefont {E.}~\bibnamefont {Kawakami}},
  \bibinfo {author} {\bibfnamefont {D.~R.}\ \bibnamefont {Ward}}, \bibinfo
  {author} {\bibfnamefont {P.}~\bibnamefont {Scarlino}}, \bibinfo {author}
  {\bibfnamefont {M.}~\bibnamefont {Veldhorst}}, \bibinfo {author}
  {\bibfnamefont {D.~E.}\ \bibnamefont {Savage}}, \bibinfo {author}
  {\bibfnamefont {M.~G.}\ \bibnamefont {Lagally}}, \bibinfo {author}
  {\bibfnamefont {M.}~\bibnamefont {Friesen}}, \bibinfo {author} {\bibfnamefont
  {S.~N.}\ \bibnamefont {Coppersmith}}, \bibinfo {author} {\bibfnamefont
  {M.~A.}\ \bibnamefont {Eriksson}}, \ and\ \bibinfo {author} {\bibfnamefont
  {L.~M.~K.}\ \bibnamefont {Vandersypen}},\ }\href {\doibase
  10.1038/nature25766} {\bibfield  {journal} {\bibinfo  {journal} {Nature}\
  }\textbf {\bibinfo {volume} {555}},\ \bibinfo {pages} {633} (\bibinfo {year}
  {2018})}\BibitemShut {NoStop}%
\bibitem [{\citenamefont {Huang}\ \emph {et~al.}(2019)\citenamefont {Huang},
  \citenamefont {Yang}, \citenamefont {Chan}, \citenamefont {Tanttu},
  \citenamefont {Hensen}, \citenamefont {Leon}, \citenamefont {Fogarty},
  \citenamefont {Hwang}, \citenamefont {Hudson}, \citenamefont {Itoh},
  \citenamefont {Morello}, \citenamefont {Laucht},\ and\ \citenamefont
  {Dzurak}}]{huang2019}%
  \BibitemOpen
  \bibfield  {author} {\bibinfo {author} {\bibfnamefont {W.}~\bibnamefont
  {Huang}}, \bibinfo {author} {\bibfnamefont {C.~H.}\ \bibnamefont {Yang}},
  \bibinfo {author} {\bibfnamefont {K.~W.}\ \bibnamefont {Chan}}, \bibinfo
  {author} {\bibfnamefont {T.}~\bibnamefont {Tanttu}}, \bibinfo {author}
  {\bibfnamefont {B.}~\bibnamefont {Hensen}}, \bibinfo {author} {\bibfnamefont
  {R.~C.~C.}\ \bibnamefont {Leon}}, \bibinfo {author} {\bibfnamefont {M.~A.}\
  \bibnamefont {Fogarty}}, \bibinfo {author} {\bibfnamefont {J.~C.~C.}\
  \bibnamefont {Hwang}}, \bibinfo {author} {\bibfnamefont {F.~E.}\ \bibnamefont
  {Hudson}}, \bibinfo {author} {\bibfnamefont {K.~M.}\ \bibnamefont {Itoh}},
  \bibinfo {author} {\bibfnamefont {A.}~\bibnamefont {Morello}}, \bibinfo
  {author} {\bibfnamefont {A.}~\bibnamefont {Laucht}}, \ and\ \bibinfo {author}
  {\bibfnamefont {A.~S.}\ \bibnamefont {Dzurak}},\ }\href {\doibase
  10.1038/s41586-019-1197-0} {\bibfield  {journal} {\bibinfo  {journal}
  {Nature}\ }\textbf {\bibinfo {volume} {569}},\ \bibinfo {pages} {532}
  (\bibinfo {year} {2019})}\BibitemShut {NoStop}%
\bibitem [{\citenamefont {Xue}\ \emph {et~al.}(2019)\citenamefont {Xue},
  \citenamefont {Watson}, \citenamefont {Helsen}, \citenamefont {Ward},
  \citenamefont {Savage}, \citenamefont {Lagally}, \citenamefont {Coppersmith},
  \citenamefont {Eriksson}, \citenamefont {Wehner},\ and\ \citenamefont
  {Vandersypen}}]{xue2019}%
  \BibitemOpen
  \bibfield  {author} {\bibinfo {author} {\bibfnamefont {X.}~\bibnamefont
  {Xue}}, \bibinfo {author} {\bibfnamefont {T.~F.}\ \bibnamefont {Watson}},
  \bibinfo {author} {\bibfnamefont {J.}~\bibnamefont {Helsen}}, \bibinfo
  {author} {\bibfnamefont {D.~R.}\ \bibnamefont {Ward}}, \bibinfo {author}
  {\bibfnamefont {D.~E.}\ \bibnamefont {Savage}}, \bibinfo {author}
  {\bibfnamefont {M.~G.}\ \bibnamefont {Lagally}}, \bibinfo {author}
  {\bibfnamefont {S.~N.}\ \bibnamefont {Coppersmith}}, \bibinfo {author}
  {\bibfnamefont {M.~A.}\ \bibnamefont {Eriksson}}, \bibinfo {author}
  {\bibfnamefont {S.}~\bibnamefont {Wehner}}, \ and\ \bibinfo {author}
  {\bibfnamefont {L.~M.~K.}\ \bibnamefont {Vandersypen}},\ }\href {\doibase
  10.1103/PhysRevX.9.021011} {\bibfield  {journal} {\bibinfo  {journal} {Phys.
  Rev. X}\ }\textbf {\bibinfo {volume} {9}},\ \bibinfo {pages} {021011}
  (\bibinfo {year} {2019})}\BibitemShut {NoStop}%
\bibitem [{\citenamefont {He}\ \emph {et~al.}(2019)\citenamefont {He},
  \citenamefont {Gorman}, \citenamefont {Keith}, \citenamefont {Kranz},
  \citenamefont {Keizer},\ and\ \citenamefont {Simmons}}]{msimmons2019}%
  \BibitemOpen
  \bibfield  {author} {\bibinfo {author} {\bibfnamefont {Y.}~\bibnamefont
  {He}}, \bibinfo {author} {\bibfnamefont {S.~K.}\ \bibnamefont {Gorman}},
  \bibinfo {author} {\bibfnamefont {D.}~\bibnamefont {Keith}}, \bibinfo
  {author} {\bibfnamefont {L.}~\bibnamefont {Kranz}}, \bibinfo {author}
  {\bibfnamefont {J.~G.}\ \bibnamefont {Keizer}}, \ and\ \bibinfo {author}
  {\bibfnamefont {M.~Y.}\ \bibnamefont {Simmons}},\ }\href {\doibase
  10.1038/s41586-019-1381-2} {\bibfield  {journal} {\bibinfo  {journal}
  {Nature}\ }\textbf {\bibinfo {volume} {571}},\ \bibinfo {pages} {371}
  (\bibinfo {year} {2019})}\BibitemShut {NoStop}%
\bibitem [{\citenamefont {Sigillito}\ \emph {et~al.}(2019)\citenamefont
  {Sigillito}, \citenamefont {Gullans}, \citenamefont {Edge}, \citenamefont
  {Borselli},\ and\ \citenamefont {Petta}}]{sigillito2019}%
  \BibitemOpen
  \bibfield  {author} {\bibinfo {author} {\bibfnamefont {A.~J.}\ \bibnamefont
  {Sigillito}}, \bibinfo {author} {\bibfnamefont {M.~J.}\ \bibnamefont
  {Gullans}}, \bibinfo {author} {\bibfnamefont {L.~F.}\ \bibnamefont {Edge}},
  \bibinfo {author} {\bibfnamefont {M.}~\bibnamefont {Borselli}}, \ and\
  \bibinfo {author} {\bibfnamefont {J.~R.}\ \bibnamefont {Petta}},\ }\href
  {\doibase 10.1038/s41534-019-0225-0} {\bibfield  {journal} {\bibinfo
  {journal} {npj Quantum Inf}\ }\textbf {\bibinfo {volume} {5}},\ \bibinfo
  {pages} {110} (\bibinfo {year} {2019})}\BibitemShut {NoStop}%
\bibitem [{\citenamefont {Borjans}\ \emph {et~al.}(2020)\citenamefont
  {Borjans}, \citenamefont {Croot}, \citenamefont {Mi}, \citenamefont
  {Gullans},\ and\ \citenamefont {Petta}}]{borjans2020}%
  \BibitemOpen
  \bibfield  {author} {\bibinfo {author} {\bibfnamefont {F.}~\bibnamefont
  {Borjans}}, \bibinfo {author} {\bibfnamefont {X.~G.}\ \bibnamefont {Croot}},
  \bibinfo {author} {\bibfnamefont {X.}~\bibnamefont {Mi}}, \bibinfo {author}
  {\bibfnamefont {M.~J.}\ \bibnamefont {Gullans}}, \ and\ \bibinfo {author}
  {\bibfnamefont {J.~R.}\ \bibnamefont {Petta}},\ }\href {\doibase
  10.1038/s41586-019-1867-y} {\bibfield  {journal} {\bibinfo  {journal}
  {Nature}\ }\textbf {\bibinfo {volume} {577}},\ \bibinfo {pages} {195–198}
  (\bibinfo {year} {2020})}\BibitemShut {NoStop}%
\bibitem [{\citenamefont {Li}\ \emph {et~al.}(2015)\citenamefont {Li},
  \citenamefont {Cao}, \citenamefont {Yu}, \citenamefont {Xiao}, \citenamefont
  {Guo}, \citenamefont {Jiang},\ and\ \citenamefont {Guo}}]{li2015}%
  \BibitemOpen
  \bibfield  {author} {\bibinfo {author} {\bibfnamefont {H.-O.}\ \bibnamefont
  {Li}}, \bibinfo {author} {\bibfnamefont {G.}~\bibnamefont {Cao}}, \bibinfo
  {author} {\bibfnamefont {G.-D.}\ \bibnamefont {Yu}}, \bibinfo {author}
  {\bibfnamefont {M.}~\bibnamefont {Xiao}}, \bibinfo {author} {\bibfnamefont
  {G.-C.}\ \bibnamefont {Guo}}, \bibinfo {author} {\bibfnamefont {H.-W.}\
  \bibnamefont {Jiang}}, \ and\ \bibinfo {author} {\bibfnamefont {G.-P.}\
  \bibnamefont {Guo}},\ }\href {\doibase 10.1038/ncomms8681} {\bibfield
  {journal} {\bibinfo  {journal} {Nat. Commun.}\ }\textbf {\bibinfo {volume}
  {6}},\ \bibinfo {pages} {8681} (\bibinfo {year} {2015})}\BibitemShut
  {NoStop}%
\bibitem [{\citenamefont {Shulman}\ \emph {et~al.}(2012)\citenamefont
  {Shulman}, \citenamefont {Dial}, \citenamefont {Harvey}, \citenamefont
  {Bluhm}, \citenamefont {Umansky},\ and\ \citenamefont
  {Yacoby}}]{shulman2012}%
  \BibitemOpen
  \bibfield  {author} {\bibinfo {author} {\bibfnamefont {M.~D.}\ \bibnamefont
  {Shulman}}, \bibinfo {author} {\bibfnamefont {O.~E.}\ \bibnamefont {Dial}},
  \bibinfo {author} {\bibfnamefont {S.~P.}\ \bibnamefont {Harvey}}, \bibinfo
  {author} {\bibfnamefont {H.}~\bibnamefont {Bluhm}}, \bibinfo {author}
  {\bibfnamefont {V.}~\bibnamefont {Umansky}}, \ and\ \bibinfo {author}
  {\bibfnamefont {A.}~\bibnamefont {Yacoby}},\ }\href {\doibase
  10.1126/science.1217692} {\bibfield  {journal} {\bibinfo  {journal}
  {Science}\ }\textbf {\bibinfo {volume} {336}},\ \bibinfo {pages} {202}
  (\bibinfo {year} {2012})}\BibitemShut {NoStop}%
\bibitem [{\citenamefont {Nichol}\ \emph {et~al.}(2017)\citenamefont {Nichol},
  \citenamefont {Orona}, \citenamefont {Harvey}, \citenamefont {Fallahi},
  \citenamefont {Gardner}, \citenamefont {Manfra},\ and\ \citenamefont
  {Yacoby}}]{nichol2017}%
  \BibitemOpen
  \bibfield  {author} {\bibinfo {author} {\bibfnamefont {J.~M.}\ \bibnamefont
  {Nichol}}, \bibinfo {author} {\bibfnamefont {L.~A.}\ \bibnamefont {Orona}},
  \bibinfo {author} {\bibfnamefont {S.~P.}\ \bibnamefont {Harvey}}, \bibinfo
  {author} {\bibfnamefont {S.}~\bibnamefont {Fallahi}}, \bibinfo {author}
  {\bibfnamefont {G.~C.}\ \bibnamefont {Gardner}}, \bibinfo {author}
  {\bibfnamefont {M.~J.}\ \bibnamefont {Manfra}}, \ and\ \bibinfo {author}
  {\bibfnamefont {A.}~\bibnamefont {Yacoby}},\ }\href {\doibase
  10.1038/s41534-016-0003-1} {\bibfield  {journal} {\bibinfo  {journal} {npj
  Quantum Inf}\ }\textbf {\bibinfo {volume} {3}},\ \bibinfo {pages} {3}
  (\bibinfo {year} {2017})}\BibitemShut {NoStop}%
\bibitem [{\citenamefont {Zajac}\ \emph {et~al.}(2016)\citenamefont {Zajac},
  \citenamefont {Hazard}, \citenamefont {Mi}, \citenamefont {Nielsen},\ and\
  \citenamefont {Petta}}]{zajac2016}%
  \BibitemOpen
  \bibfield  {author} {\bibinfo {author} {\bibfnamefont {D.~M.}\ \bibnamefont
  {Zajac}}, \bibinfo {author} {\bibfnamefont {T.~M.}\ \bibnamefont {Hazard}},
  \bibinfo {author} {\bibfnamefont {X.}~\bibnamefont {Mi}}, \bibinfo {author}
  {\bibfnamefont {E.}~\bibnamefont {Nielsen}}, \ and\ \bibinfo {author}
  {\bibfnamefont {J.~R.}\ \bibnamefont {Petta}},\ }\href {\doibase
  10.1103/PhysRevApplied.6.054013} {\bibfield  {journal} {\bibinfo  {journal}
  {Phys. Rev. Applied}\ }\textbf {\bibinfo {volume} {6}},\ \bibinfo {pages}
  {054013} (\bibinfo {year} {2016})}\BibitemShut {NoStop}%
\bibitem [{\citenamefont {Neyens}\ \emph {et~al.}(2019)\citenamefont {Neyens},
  \citenamefont {MacQuarrie}, \citenamefont {Dodson}, \citenamefont {Corrigan},
  \citenamefont {Holman}, \citenamefont {Thorgrimsson}, \citenamefont {Palma},
  \citenamefont {McJunkin}, \citenamefont {Edge}, \citenamefont {Friesen},
  \citenamefont {Coppersmith},\ and\ \citenamefont {Eriksson}}]{neyens2019}%
  \BibitemOpen
  \bibfield  {author} {\bibinfo {author} {\bibfnamefont {S.~F.}\ \bibnamefont
  {Neyens}}, \bibinfo {author} {\bibfnamefont {E.~R.}\ \bibnamefont
  {MacQuarrie}}, \bibinfo {author} {\bibfnamefont {J.~P.}\ \bibnamefont
  {Dodson}}, \bibinfo {author} {\bibfnamefont {J.}~\bibnamefont {Corrigan}},
  \bibinfo {author} {\bibfnamefont {N.}~\bibnamefont {Holman}}, \bibinfo
  {author} {\bibfnamefont {B.}~\bibnamefont {Thorgrimsson}}, \bibinfo {author}
  {\bibfnamefont {M.}~\bibnamefont {Palma}}, \bibinfo {author} {\bibfnamefont
  {T.}~\bibnamefont {McJunkin}}, \bibinfo {author} {\bibfnamefont {L.~F.}\
  \bibnamefont {Edge}}, \bibinfo {author} {\bibfnamefont {M.}~\bibnamefont
  {Friesen}}, \bibinfo {author} {\bibfnamefont {S.~N.}\ \bibnamefont
  {Coppersmith}}, \ and\ \bibinfo {author} {\bibfnamefont {M.~A.}\ \bibnamefont
  {Eriksson}},\ }\href {\doibase 10.1103/PhysRevApplied.12.064049} {\bibfield
  {journal} {\bibinfo  {journal} {Phys. Rev. Applied}\ }\textbf {\bibinfo
  {volume} {12}} (\bibinfo {year} {2019}),\
  10.1103/PhysRevApplied.12.064049}\BibitemShut {NoStop}%
\bibitem [{\citenamefont {Ward}\ \emph {et~al.}(2016)\citenamefont {Ward},
  \citenamefont {Kim}, \citenamefont {Savage}, \citenamefont {Lagally},
  \citenamefont {Foote}, \citenamefont {Friesen}, \citenamefont {Coppersmith},\
  and\ \citenamefont {Eriksson}}]{ward2016}%
  \BibitemOpen
  \bibfield  {author} {\bibinfo {author} {\bibfnamefont {D.~R.}\ \bibnamefont
  {Ward}}, \bibinfo {author} {\bibfnamefont {D.}~\bibnamefont {Kim}}, \bibinfo
  {author} {\bibfnamefont {D.~E.}\ \bibnamefont {Savage}}, \bibinfo {author}
  {\bibfnamefont {M.~G.}\ \bibnamefont {Lagally}}, \bibinfo {author}
  {\bibfnamefont {R.~H.}\ \bibnamefont {Foote}}, \bibinfo {author}
  {\bibfnamefont {M.}~\bibnamefont {Friesen}}, \bibinfo {author} {\bibfnamefont
  {S.~N.}\ \bibnamefont {Coppersmith}}, \ and\ \bibinfo {author} {\bibfnamefont
  {M.~A.}\ \bibnamefont {Eriksson}},\ }\href {\doibase 10.1038/npjqi.2016.32}
  {\bibfield  {journal} {\bibinfo  {journal} {npj Quantum Inf}\ }\textbf
  {\bibinfo {volume} {2}},\ \bibinfo {pages} {16032} (\bibinfo {year}
  {2016})}\BibitemShut {NoStop}%
\bibitem [{\citenamefont {Angus}\ \emph {et~al.}(2007)\citenamefont {Angus},
  \citenamefont {Ferguson}, \citenamefont {Dzurak},\ and\ \citenamefont
  {Clark}}]{angus2007}%
  \BibitemOpen
  \bibfield  {author} {\bibinfo {author} {\bibfnamefont {S.~J.}\ \bibnamefont
  {Angus}}, \bibinfo {author} {\bibfnamefont {A.~J.}\ \bibnamefont {Ferguson}},
  \bibinfo {author} {\bibfnamefont {A.~S.}\ \bibnamefont {Dzurak}}, \ and\
  \bibinfo {author} {\bibfnamefont {R.~G.}\ \bibnamefont {Clark}},\ }\href
  {\doibase 10.1021/nl070949k} {\bibfield  {journal} {\bibinfo  {journal} {Nano
  Lett.}\ }\textbf {\bibinfo {volume} {7}},\ \bibinfo {pages} {2051} (\bibinfo
  {year} {2007})}\BibitemShut {NoStop}%
\bibitem [{\citenamefont {Borselli}\ \emph {et~al.}(2015)\citenamefont
  {Borselli}, \citenamefont {Eng}, \citenamefont {Ross}, \citenamefont
  {Hazard}, \citenamefont {Holabird}, \citenamefont {Huang}, \citenamefont
  {Kiselev}, \citenamefont {Deelman}, \citenamefont {Warren}, \citenamefont
  {Milosavljevic}, \citenamefont {Schmitz}, \citenamefont {Sokolich},
  \citenamefont {Gyure},\ and\ \citenamefont {Hunter}}]{borselli2015}%
  \BibitemOpen
  \bibfield  {author} {\bibinfo {author} {\bibfnamefont {M.~G.}\ \bibnamefont
  {Borselli}}, \bibinfo {author} {\bibfnamefont {K.}~\bibnamefont {Eng}},
  \bibinfo {author} {\bibfnamefont {R.~S.}\ \bibnamefont {Ross}}, \bibinfo
  {author} {\bibfnamefont {T.~M.}\ \bibnamefont {Hazard}}, \bibinfo {author}
  {\bibfnamefont {K.~S.}\ \bibnamefont {Holabird}}, \bibinfo {author}
  {\bibfnamefont {B.}~\bibnamefont {Huang}}, \bibinfo {author} {\bibfnamefont
  {A.~A.}\ \bibnamefont {Kiselev}}, \bibinfo {author} {\bibfnamefont {P.~W.}\
  \bibnamefont {Deelman}}, \bibinfo {author} {\bibfnamefont {L.~D.}\
  \bibnamefont {Warren}}, \bibinfo {author} {\bibfnamefont {I.}~\bibnamefont
  {Milosavljevic}}, \bibinfo {author} {\bibfnamefont {A.~E.}\ \bibnamefont
  {Schmitz}}, \bibinfo {author} {\bibfnamefont {M.}~\bibnamefont {Sokolich}},
  \bibinfo {author} {\bibfnamefont {M.~F.}\ \bibnamefont {Gyure}}, \ and\
  \bibinfo {author} {\bibfnamefont {A.~T.}\ \bibnamefont {Hunter}},\ }\href
  {\doibase 10.1088/0957-4484/26/37/375202} {\bibfield  {journal} {\bibinfo
  {journal} {Nanotechnology}\ }\textbf {\bibinfo {volume} {26}},\ \bibinfo
  {pages} {375202} (\bibinfo {year} {2015})}\BibitemShut {NoStop}%
\bibitem [{SI()}]{SI}%
  \BibitemOpen
  \href@noop {} {}\bibinfo {note} {See Supplementary Information}\BibitemShut
  {NoStop}%
\bibitem [{\citenamefont {Tracy}\ \emph {et~al.}(2016)\citenamefont {Tracy},
  \citenamefont {Luhman}, \citenamefont {Carr}, \citenamefont {Bishop},
  \citenamefont {Ten~Eyck}, \citenamefont {Pluym}, \citenamefont {Wendt},
  \citenamefont {Lilly},\ and\ \citenamefont {Carroll}}]{tracy2016}%
  \BibitemOpen
  \bibfield  {author} {\bibinfo {author} {\bibfnamefont {L.~A.}\ \bibnamefont
  {Tracy}}, \bibinfo {author} {\bibfnamefont {D.~R.}\ \bibnamefont {Luhman}},
  \bibinfo {author} {\bibfnamefont {S.~M.}\ \bibnamefont {Carr}}, \bibinfo
  {author} {\bibfnamefont {N.~C.}\ \bibnamefont {Bishop}}, \bibinfo {author}
  {\bibfnamefont {G.~A.}\ \bibnamefont {Ten~Eyck}}, \bibinfo {author}
  {\bibfnamefont {T.}~\bibnamefont {Pluym}}, \bibinfo {author} {\bibfnamefont
  {J.~R.}\ \bibnamefont {Wendt}}, \bibinfo {author} {\bibfnamefont {M.~P.}\
  \bibnamefont {Lilly}}, \ and\ \bibinfo {author} {\bibfnamefont {M.~S.}\
  \bibnamefont {Carroll}},\ }\href {\doibase 10.1063/1.4941421} {\bibfield
  {journal} {\bibinfo  {journal} {Appl. Phys. Lett.}\ }\textbf {\bibinfo
  {volume} {108}},\ \bibinfo {pages} {063101} (\bibinfo {year}
  {2016})}\BibitemShut {NoStop}%
\bibitem [{\citenamefont {Studenikin}\ \emph {et~al.}(2012)\citenamefont
  {Studenikin}, \citenamefont {Thorgrimson}, \citenamefont {Aers},
  \citenamefont {Kam}, \citenamefont {Zawadzki}, \citenamefont {Wasilewski},
  \citenamefont {Bogan},\ and\ \citenamefont {Sachrajda}}]{studenikin2012}%
  \BibitemOpen
  \bibfield  {author} {\bibinfo {author} {\bibfnamefont {S.~A.}\ \bibnamefont
  {Studenikin}}, \bibinfo {author} {\bibfnamefont {J.}~\bibnamefont
  {Thorgrimson}}, \bibinfo {author} {\bibfnamefont {G.~C.}\ \bibnamefont
  {Aers}}, \bibinfo {author} {\bibfnamefont {A.}~\bibnamefont {Kam}}, \bibinfo
  {author} {\bibfnamefont {P.}~\bibnamefont {Zawadzki}}, \bibinfo {author}
  {\bibfnamefont {Z.~R.}\ \bibnamefont {Wasilewski}}, \bibinfo {author}
  {\bibfnamefont {A.}~\bibnamefont {Bogan}}, \ and\ \bibinfo {author}
  {\bibfnamefont {A.~S.}\ \bibnamefont {Sachrajda}},\ }\href {\doibase
  10.1063/1.4749281} {\bibfield  {journal} {\bibinfo  {journal} {Appl. Phys.
  Lett.}\ }\textbf {\bibinfo {volume} {101}},\ \bibinfo {pages} {233101}
  (\bibinfo {year} {2012})}\BibitemShut {NoStop}%
\bibitem [{\citenamefont {Harvey-Collard}\ \emph {et~al.}(2018)\citenamefont
  {Harvey-Collard}, \citenamefont {D'Anjou}, \citenamefont {Rudolph},
  \citenamefont {Jacobson}, \citenamefont {Dominguez}, \citenamefont
  {Ten~Eyck}, \citenamefont {Wendt}, \citenamefont {Pluym}, \citenamefont
  {Lilly}, \citenamefont {Coish}, \citenamefont {Pioro-Ladri\`ere},\ and\
  \citenamefont {Carroll}}]{harveycollard2018}%
  \BibitemOpen
  \bibfield  {author} {\bibinfo {author} {\bibfnamefont {P.}~\bibnamefont
  {Harvey-Collard}}, \bibinfo {author} {\bibfnamefont {B.}~\bibnamefont
  {D'Anjou}}, \bibinfo {author} {\bibfnamefont {M.}~\bibnamefont {Rudolph}},
  \bibinfo {author} {\bibfnamefont {N.~T.}\ \bibnamefont {Jacobson}}, \bibinfo
  {author} {\bibfnamefont {J.}~\bibnamefont {Dominguez}}, \bibinfo {author}
  {\bibfnamefont {G.~A.}\ \bibnamefont {Ten~Eyck}}, \bibinfo {author}
  {\bibfnamefont {J.~R.}\ \bibnamefont {Wendt}}, \bibinfo {author}
  {\bibfnamefont {T.}~\bibnamefont {Pluym}}, \bibinfo {author} {\bibfnamefont
  {M.~P.}\ \bibnamefont {Lilly}}, \bibinfo {author} {\bibfnamefont {W.~A.}\
  \bibnamefont {Coish}}, \bibinfo {author} {\bibfnamefont {M.}~\bibnamefont
  {Pioro-Ladri\`ere}}, \ and\ \bibinfo {author} {\bibfnamefont {M.~S.}\
  \bibnamefont {Carroll}},\ }\href {\doibase 10.1103/PhysRevX.8.021046}
  {\bibfield  {journal} {\bibinfo  {journal} {Phys. Rev. X}\ }\textbf {\bibinfo
  {volume} {8}},\ \bibinfo {pages} {021046} (\bibinfo {year}
  {2018})}\BibitemShut {NoStop}%
\bibitem [{\citenamefont {Fujisawa}\ \emph {et~al.}(2004)\citenamefont
  {Fujisawa}, \citenamefont {Hayashi}, \citenamefont {Cheong}, \citenamefont
  {Jeong},\ and\ \citenamefont {Hirayama}}]{fujisawa2004}%
  \BibitemOpen
  \bibfield  {author} {\bibinfo {author} {\bibfnamefont {T.}~\bibnamefont
  {Fujisawa}}, \bibinfo {author} {\bibfnamefont {T.}~\bibnamefont {Hayashi}},
  \bibinfo {author} {\bibfnamefont {H.}~\bibnamefont {Cheong}}, \bibinfo
  {author} {\bibfnamefont {Y.}~\bibnamefont {Jeong}}, \ and\ \bibinfo {author}
  {\bibfnamefont {Y.}~\bibnamefont {Hirayama}},\ }\href {\doibase
  https://doi.org/10.1016/j.physe.2003.11.184} {\bibfield  {journal} {\bibinfo
  {journal} {Physica E: Low-dimensional Systems and Nanostructures}\ }\textbf
  {\bibinfo {volume} {21}},\ \bibinfo {pages} {1046 } (\bibinfo {year}
  {2004})}\BibitemShut {NoStop}%
\bibitem [{\citenamefont {Mi}\ \emph {et~al.}(2018)\citenamefont {Mi},
  \citenamefont {Kohler},\ and\ \citenamefont {Petta}}]{miValley2018}%
  \BibitemOpen
  \bibfield  {author} {\bibinfo {author} {\bibfnamefont {X.}~\bibnamefont
  {Mi}}, \bibinfo {author} {\bibfnamefont {S.}~\bibnamefont {Kohler}}, \ and\
  \bibinfo {author} {\bibfnamefont {J.~R.}\ \bibnamefont {Petta}},\ }\href
  {\doibase 10.1103/PhysRevB.98.161404} {\bibfield  {journal} {\bibinfo
  {journal} {Phys. Rev. B}\ }\textbf {\bibinfo {volume} {98}},\ \bibinfo
  {pages} {161404} (\bibinfo {year} {2018})}\BibitemShut {NoStop}%
\bibitem [{\citenamefont {Dial}\ \emph {et~al.}(2013)\citenamefont {Dial},
  \citenamefont {Shulman}, \citenamefont {Harvey}, \citenamefont {Bluhm},
  \citenamefont {Umansky},\ and\ \citenamefont {Yacoby}}]{dial2013}%
  \BibitemOpen
  \bibfield  {author} {\bibinfo {author} {\bibfnamefont {O.~E.}\ \bibnamefont
  {Dial}}, \bibinfo {author} {\bibfnamefont {M.~D.}\ \bibnamefont {Shulman}},
  \bibinfo {author} {\bibfnamefont {S.~P.}\ \bibnamefont {Harvey}}, \bibinfo
  {author} {\bibfnamefont {H.}~\bibnamefont {Bluhm}}, \bibinfo {author}
  {\bibfnamefont {V.}~\bibnamefont {Umansky}}, \ and\ \bibinfo {author}
  {\bibfnamefont {A.}~\bibnamefont {Yacoby}},\ }\href {\doibase
  10.1103/PhysRevLett.110.146804} {\bibfield  {journal} {\bibinfo  {journal}
  {Phys. Rev. Lett.}\ }\textbf {\bibinfo {volume} {110}},\ \bibinfo {pages}
  {146804} (\bibinfo {year} {2013})}\BibitemShut {NoStop}%
\bibitem [{\citenamefont {Frees}\ \emph {et~al.}(2019)\citenamefont {Frees},
  \citenamefont {Mehl}, \citenamefont {Gamble}, \citenamefont {Friesen},\ and\
  \citenamefont {Coppersmith}}]{frees2018}%
  \BibitemOpen
  \bibfield  {author} {\bibinfo {author} {\bibfnamefont {A.}~\bibnamefont
  {Frees}}, \bibinfo {author} {\bibfnamefont {S.}~\bibnamefont {Mehl}},
  \bibinfo {author} {\bibfnamefont {J.~K.}\ \bibnamefont {Gamble}}, \bibinfo
  {author} {\bibfnamefont {M.}~\bibnamefont {Friesen}}, \ and\ \bibinfo
  {author} {\bibfnamefont {S.~N.}\ \bibnamefont {Coppersmith}},\ }\href
  {\doibase 10.1038/s41534-019-0190-7} {\bibfield  {journal} {\bibinfo
  {journal} {npj Quantum Inf}\ }\textbf {\bibinfo {volume} {5}},\ \bibinfo
  {pages} {73} (\bibinfo {year} {2019})}\BibitemShut {NoStop}%
\bibitem [{\citenamefont {Vandersypen}\ and\ \citenamefont
  {Chuang}(2005)}]{vandersypen2005}%
  \BibitemOpen
  \bibfield  {author} {\bibinfo {author} {\bibfnamefont {L.~M.~K.}\
  \bibnamefont {Vandersypen}}\ and\ \bibinfo {author} {\bibfnamefont {I.~L.}\
  \bibnamefont {Chuang}},\ }\href {\doibase 10.1103/RevModPhys.76.1037}
  {\bibfield  {journal} {\bibinfo  {journal} {Rev. Mod. Phys.}\ }\textbf
  {\bibinfo {volume} {76}},\ \bibinfo {pages} {1037} (\bibinfo {year}
  {2005})}\BibitemShut {NoStop}%
\bibitem [{\citenamefont {James}\ \emph {et~al.}(2001)\citenamefont {James},
  \citenamefont {Kwiat}, \citenamefont {Munro},\ and\ \citenamefont
  {White}}]{james2001}%
  \BibitemOpen
  \bibfield  {author} {\bibinfo {author} {\bibfnamefont {D.~F.~V.}\
  \bibnamefont {James}}, \bibinfo {author} {\bibfnamefont {P.~G.}\ \bibnamefont
  {Kwiat}}, \bibinfo {author} {\bibfnamefont {W.~J.}\ \bibnamefont {Munro}}, \
  and\ \bibinfo {author} {\bibfnamefont {A.~G.}\ \bibnamefont {White}},\ }\href
  {\doibase 10.1103/PhysRevA.64.052312} {\bibfield  {journal} {\bibinfo
  {journal} {Phys. Rev. A}\ }\textbf {\bibinfo {volume} {64}},\ \bibinfo
  {pages} {052312} (\bibinfo {year} {2001})}\BibitemShut {NoStop}%
\bibitem [{\citenamefont {White}\ \emph {et~al.}(2007)\citenamefont {White},
  \citenamefont {Gilchrist}, \citenamefont {Pryde}, \citenamefont {O'Brien},
  \citenamefont {Bremner},\ and\ \citenamefont {Langford}}]{white2007}%
  \BibitemOpen
  \bibfield  {author} {\bibinfo {author} {\bibfnamefont {A.~G.}\ \bibnamefont
  {White}}, \bibinfo {author} {\bibfnamefont {A.}~\bibnamefont {Gilchrist}},
  \bibinfo {author} {\bibfnamefont {G.~J.}\ \bibnamefont {Pryde}}, \bibinfo
  {author} {\bibfnamefont {J.~L.}\ \bibnamefont {O'Brien}}, \bibinfo {author}
  {\bibfnamefont {M.~J.}\ \bibnamefont {Bremner}}, \ and\ \bibinfo {author}
  {\bibfnamefont {N.~K.}\ \bibnamefont {Langford}},\ }\href {\doibase
  10.1364/JOSAB.24.000172} {\bibfield  {journal} {\bibinfo  {journal} {J. Opt.
  Soc. Am. B}\ }\textbf {\bibinfo {volume} {24}},\ \bibinfo {pages} {172}
  (\bibinfo {year} {2007})}\BibitemShut {NoStop}%
\bibitem [{\citenamefont {Kim}\ \emph {et~al.}(2015{\natexlab{a}})\citenamefont
  {Kim}, \citenamefont {Ward}, \citenamefont {Simmons}, \citenamefont {Savage},
  \citenamefont {Lagally}, \citenamefont {Friesen}, \citenamefont
  {Coppersmith},\ and\ \citenamefont {Eriksson}}]{kim2015}%
  \BibitemOpen
  \bibfield  {author} {\bibinfo {author} {\bibfnamefont {D.}~\bibnamefont
  {Kim}}, \bibinfo {author} {\bibfnamefont {D.~R.}\ \bibnamefont {Ward}},
  \bibinfo {author} {\bibfnamefont {C.~B.}\ \bibnamefont {Simmons}}, \bibinfo
  {author} {\bibfnamefont {D.~E.}\ \bibnamefont {Savage}}, \bibinfo {author}
  {\bibfnamefont {M.~G.}\ \bibnamefont {Lagally}}, \bibinfo {author}
  {\bibfnamefont {M.}~\bibnamefont {Friesen}}, \bibinfo {author} {\bibfnamefont
  {S.~N.}\ \bibnamefont {Coppersmith}}, \ and\ \bibinfo {author} {\bibfnamefont
  {M.~A.}\ \bibnamefont {Eriksson}},\ }\href {\doibase 10.1038/npjqi.2015.4}
  {\bibfield  {journal} {\bibinfo  {journal} {npj Quantum Inf}\ }\textbf
  {\bibinfo {volume} {1}},\ \bibinfo {pages} {15004} (\bibinfo {year}
  {2015}{\natexlab{a}})}\BibitemShut {NoStop}%
\bibitem [{\citenamefont {Thorgrimsson}\ \emph {et~al.}(2017)\citenamefont
  {Thorgrimsson}, \citenamefont {Kim}, \citenamefont {Yang}, \citenamefont
  {Smith}, \citenamefont {Simmons}, \citenamefont {Ward}, \citenamefont
  {Foote}, \citenamefont {Corrigan}, \citenamefont {Savage}, \citenamefont
  {Lagally}, \citenamefont {Friesen}, \citenamefont {Coppersmith},\ and\
  \citenamefont {Eriksson}}]{thorgrimsson2017}%
  \BibitemOpen
  \bibfield  {author} {\bibinfo {author} {\bibfnamefont {B.}~\bibnamefont
  {Thorgrimsson}}, \bibinfo {author} {\bibfnamefont {D.}~\bibnamefont {Kim}},
  \bibinfo {author} {\bibfnamefont {Y.-C.}\ \bibnamefont {Yang}}, \bibinfo
  {author} {\bibfnamefont {L.~W.}\ \bibnamefont {Smith}}, \bibinfo {author}
  {\bibfnamefont {C.~B.}\ \bibnamefont {Simmons}}, \bibinfo {author}
  {\bibfnamefont {D.~R.}\ \bibnamefont {Ward}}, \bibinfo {author}
  {\bibfnamefont {R.~H.}\ \bibnamefont {Foote}}, \bibinfo {author}
  {\bibfnamefont {J.}~\bibnamefont {Corrigan}}, \bibinfo {author}
  {\bibfnamefont {D.~E.}\ \bibnamefont {Savage}}, \bibinfo {author}
  {\bibfnamefont {M.~G.}\ \bibnamefont {Lagally}}, \bibinfo {author}
  {\bibfnamefont {M.}~\bibnamefont {Friesen}}, \bibinfo {author} {\bibfnamefont
  {S.~N.}\ \bibnamefont {Coppersmith}}, \ and\ \bibinfo {author} {\bibfnamefont
  {M.~A.}\ \bibnamefont {Eriksson}},\ }\href {\doibase
  10.1038/s41534-017-0034-2} {\bibfield  {journal} {\bibinfo  {journal} {npj
  Quantum Inf}\ }\textbf {\bibinfo {volume} {3}},\ \bibinfo {pages} {32}
  (\bibinfo {year} {2017})}\BibitemShut {NoStop}%
\bibitem [{\citenamefont {DiCarlo}\ \emph {et~al.}(2004)\citenamefont
  {DiCarlo}, \citenamefont {Lynch}, \citenamefont {Johnson}, \citenamefont
  {Childress}, \citenamefont {Crockett}, \citenamefont {Marcus}, \citenamefont
  {Hanson},\ and\ \citenamefont {Gossard}}]{dicarlo2004}%
  \BibitemOpen
  \bibfield  {author} {\bibinfo {author} {\bibfnamefont {L.}~\bibnamefont
  {DiCarlo}}, \bibinfo {author} {\bibfnamefont {H.~J.}\ \bibnamefont {Lynch}},
  \bibinfo {author} {\bibfnamefont {A.~C.}\ \bibnamefont {Johnson}}, \bibinfo
  {author} {\bibfnamefont {L.~I.}\ \bibnamefont {Childress}}, \bibinfo {author}
  {\bibfnamefont {K.}~\bibnamefont {Crockett}}, \bibinfo {author}
  {\bibfnamefont {C.~M.}\ \bibnamefont {Marcus}}, \bibinfo {author}
  {\bibfnamefont {M.~P.}\ \bibnamefont {Hanson}}, \ and\ \bibinfo {author}
  {\bibfnamefont {A.~C.}\ \bibnamefont {Gossard}},\ }\href {\doibase
  10.1103/PhysRevLett.92.226801} {\bibfield  {journal} {\bibinfo  {journal}
  {Phys. Rev. Lett.}\ }\textbf {\bibinfo {volume} {92}},\ \bibinfo {pages}
  {226801} (\bibinfo {year} {2004})}\BibitemShut {NoStop}%
\bibitem [{\citenamefont {Zhao}\ and\ \citenamefont {Hu}(2018)}]{hu2018}%
  \BibitemOpen
  \bibfield  {author} {\bibinfo {author} {\bibfnamefont {X.}~\bibnamefont
  {Zhao}}\ and\ \bibinfo {author} {\bibfnamefont {X.}~\bibnamefont {Hu}},\
  }\href {http://arxiv.org/abs/1803.00749} {\bibfield  {journal} {\bibinfo
  {journal} {arXiv:1803.00749}\ } (\bibinfo {year} {2018})}\BibitemShut
  {NoStop}%
\bibitem [{\citenamefont {Maradan}\ \emph {et~al.}(2014)\citenamefont
  {Maradan}, \citenamefont {Casparis}, \citenamefont {Liu}, \citenamefont
  {Biesinger}, \citenamefont {Scheller}, \citenamefont {Zumb{\"u}hl},
  \citenamefont {Zimmerman},\ and\ \citenamefont {Gossard}}]{maradan2014}%
  \BibitemOpen
  \bibfield  {author} {\bibinfo {author} {\bibfnamefont {D.}~\bibnamefont
  {Maradan}}, \bibinfo {author} {\bibfnamefont {L.}~\bibnamefont {Casparis}},
  \bibinfo {author} {\bibfnamefont {T.-M.}\ \bibnamefont {Liu}}, \bibinfo
  {author} {\bibfnamefont {D.~E.~F.}\ \bibnamefont {Biesinger}}, \bibinfo
  {author} {\bibfnamefont {C.~P.}\ \bibnamefont {Scheller}}, \bibinfo {author}
  {\bibfnamefont {D.~M.}\ \bibnamefont {Zumb{\"u}hl}}, \bibinfo {author}
  {\bibfnamefont {J.~D.}\ \bibnamefont {Zimmerman}}, \ and\ \bibinfo {author}
  {\bibfnamefont {A.~C.}\ \bibnamefont {Gossard}},\ }\href {\doibase
  10.1007/s10909-014-1169-6} {\bibfield  {journal} {\bibinfo  {journal}
  {Journal of Low Temperature Physics}\ }\textbf {\bibinfo {volume} {175}},\
  \bibinfo {pages} {784} (\bibinfo {year} {2014})}\BibitemShut {NoStop}%
\bibitem [{\citenamefont {Wang}\ \emph {et~al.}(2013)\citenamefont {Wang},
  \citenamefont {Payette}, \citenamefont {Dovzhenko}, \citenamefont {Deelman},\
  and\ \citenamefont {Petta}}]{wang2013}%
  \BibitemOpen
  \bibfield  {author} {\bibinfo {author} {\bibfnamefont {K.}~\bibnamefont
  {Wang}}, \bibinfo {author} {\bibfnamefont {C.}~\bibnamefont {Payette}},
  \bibinfo {author} {\bibfnamefont {Y.}~\bibnamefont {Dovzhenko}}, \bibinfo
  {author} {\bibfnamefont {P.~W.}\ \bibnamefont {Deelman}}, \ and\ \bibinfo
  {author} {\bibfnamefont {J.~R.}\ \bibnamefont {Petta}},\ }\href {\doibase
  10.1103/PhysRevLett.111.046801} {\bibfield  {journal} {\bibinfo  {journal}
  {Phys. Rev. Lett.}\ }\textbf {\bibinfo {volume} {111}},\ \bibinfo {pages}
  {046801} (\bibinfo {year} {2013})}\BibitemShut {NoStop}%
\bibitem [{\citenamefont {Kim}\ \emph {et~al.}(2015{\natexlab{b}})\citenamefont
  {Kim}, \citenamefont {Ward}, \citenamefont {Simmons}, \citenamefont {Gamble},
  \citenamefont {Blume-Kohout}, \citenamefont {Nielsen}, \citenamefont
  {Savage}, \citenamefont {Lagally}, \citenamefont {Friesen}, \citenamefont
  {Coppersmith},\ and\ \citenamefont {Eriksson}}]{dohun2015}%
  \BibitemOpen
  \bibfield  {author} {\bibinfo {author} {\bibfnamefont {D.}~\bibnamefont
  {Kim}}, \bibinfo {author} {\bibfnamefont {D.~R.}\ \bibnamefont {Ward}},
  \bibinfo {author} {\bibfnamefont {C.~B.}\ \bibnamefont {Simmons}}, \bibinfo
  {author} {\bibfnamefont {J.~K.}\ \bibnamefont {Gamble}}, \bibinfo {author}
  {\bibfnamefont {R.}~\bibnamefont {Blume-Kohout}}, \bibinfo {author}
  {\bibfnamefont {E.}~\bibnamefont {Nielsen}}, \bibinfo {author} {\bibfnamefont
  {D.~E.}\ \bibnamefont {Savage}}, \bibinfo {author} {\bibfnamefont {M.~G.}\
  \bibnamefont {Lagally}}, \bibinfo {author} {\bibfnamefont {M.}~\bibnamefont
  {Friesen}}, \bibinfo {author} {\bibfnamefont {S.~N.}\ \bibnamefont
  {Coppersmith}}, \ and\ \bibinfo {author} {\bibfnamefont {M.~A.}\ \bibnamefont
  {Eriksson}},\ }\href {\doibase 10.1038/nnano.2014.336} {\bibfield  {journal}
  {\bibinfo  {journal} {Nature Nanotech.}\ }\textbf {\bibinfo {volume} {10}},\
  \bibinfo {pages} {243} (\bibinfo {year} {2015}{\natexlab{b}})}\BibitemShut
  {NoStop}%
\end{thebibliography}%


%

\end{document}